\documentclass[letterpaper,twocolumn,10pt]{article}
\usepackage{usenix2019_v3}

\usepackage[compact]{titlesec}

\usepackage{tikz}
\usepackage{amsmath}
\usepackage{courier}
\usepackage{xspace}
\usepackage{microtype}
\usepackage[normalem]{ulem}
\usepackage{subfig}
\usepackage{enumitem}
\usepackage[super]{nth}
\usepackage{cite}
\usepackage{xcolor}
\newcommand{\eg}{\emph{e.g.}\xspace}

\newcommand{\ie}{\emph{i.e.}\xspace}
\newcommand{\etal}{\emph{et al.}\xspace}
\newcommand{\myparagraph}[1]{\noindent\textbf{#1.}}

\makeatletter%
\begin{document}

\title{
  Serverless in the Wild: Characterizing and Optimizing \\
  the Serverless Workload at a Large Cloud Provider
  \vspace*{-.05in}
}
\author{{\rm Mohammad Shahrad, Rodrigo Fonseca, \'I\~{n}igo Goiri,
  Gohar Chaudhry, Paul Batum,} \\ {\rm Jason Cooke, Eduardo Laureano,
  Colby Tresness, Mark Russinovich, and Ricardo Bianchini}
  \thanks{Shahrad is affiliated with Princeton University,
    but was at MSR during this work. Laureano
    and Tresness are now with Facebook and D. E. Shaw.} \\
  \\
  {\rm Microsoft Azure and Microsoft Research}}
\date{}
\maketitle

\thispagestyle{empty}
\pagestyle{plain}

\begin{abstract}
Function as a Service (FaaS) has been gaining popularity as a way to
deploy computations to serverless backends in the cloud.
This paradigm shifts the complexity of allocating and provisioning
resources to the cloud provider, which has to provide the illusion of
always-available resources (\ie, fast function invocations without
cold starts) at the lowest possible resource cost. Doing so requires
the provider to deeply understand the characteristics of the FaaS
workload. Unfortunately, there has been little to no public
information on these characteristics. Thus, in this paper, we first
characterize the entire production FaaS workload
of Azure Functions.
We show for example that most functions are invoked very
infrequently, but there is an 8-order-of-magnitude range of invocation
frequencies. Using observations from our characterization, we then
propose a practical resource management policy that significantly
reduces the number of function cold starts, while spending fewer
resources than state-of-the-practice policies.
\end{abstract}

\section{Introduction}

Function as a Service (FaaS) is a software paradigm that is becoming
increasingly popular.  Multiple cloud providers offer
FaaS~\cite{azurefunctions,gcp,awslambda,ibmfunctions} as the interface
to usage-driven, stateless (serverless) backend services.  FaaS offers
an intuitive, event-based interface for developing cloud-based
applications.  In contrast with the traditional cloud interface, in
FaaS, users do not explicitly provision or configure virtual machines
(VMs) or containers.  FaaS users do not pay for resources they do not
use either.  Instead, users simply upload the code of their functions
to the cloud; functions get executed when ``triggered'' or ``invoked''
by events, such as the receipt of a message (\eg, an HTTP request) or
a timer going off.  The provider is then responsible for provisioning
the needed resources (\eg, a container in which to execute each
function), providing high function performance, and billing users just
for their actual function executions (\eg, in increments of 100
milliseconds).

Obviously, providers seek to achieve high function performance at the
lowest possible resource cost.  There are three main aspects to how
fast functions can execute and the resources they consume.  First,
function execution requires having the needed code (\eg, user code,
language runtime libraries) in memory.  A function can be started
quickly when the code is already in memory (warm start) and does not
have to be brought in from persistent storage (cold start).  Second,
keeping the resources required by all functions in memory at all times
may be prohibitively expensive for the provider, especially if
function executions are short and infrequent.  Ideally, the provider
wants to give the illusion that all functions are always warm, while
spending resources as if they were always cold.  Third, functions may
have widely varying resource needs and invocation frequencies from
multiple triggers.  These characteristics severely complicate any
attempts to predict invocations for reducing resource usage.  For
example, the wide range of invocation frequencies suggests that
keeping resources in memory may work well for some functions but not
others.  With respect to triggers, HTTP triggers may produce
invocations at irregular intervals that are difficult to predict,
whereas timers are regular.

These observations make it clear that providing high function
performance at low cost requires a deep understanding of the
characteristics of the FaaS workload.  Unfortunately, there has been
no public information on the characteristics of production
workloads.  Prior work~\cite{akkus2018sand, Figiela_2018,
Kuhlenkamp_sac_2020, Lee_2018, McGrath_2017, Wang_2018} has focused
on either (1) running benchmark functions to assess performance
and/or reverse-engineer how providers manage resources; or (2)
implementing prototype systems to run benchmark functions.  In
contrast, what is needed is a comprehensive characterization of the
users' {\em real} FaaS workloads on a {\em production} platform from
the provider's perspective.

\myparagraph{Characterizing production workloads} To fill this gap, in
this paper, we first characterize the entire production FaaS workload
of Azure Functions~\cite{azurefunctions}.
We characterize the real functions and
their trigger types, invocation frequencies and patterns, and resource
needs.  The characterization produces many interesting observations.
For example, it shows that most functions are invoked very
infrequently, but the most popular functions are invoked 8 orders of
magnitude more frequently than the least popular ones. It also shows
that functions exhibit a variety of triggers, producing invocation
patterns that are often difficult to predict.  In terms of resource
needs, the characterization shows a 4x range of function memory usage
and that 50\% of functions run in less than 1 second.

Researchers can use the distributions of the workload characteristics
we study to create realistic traces for their work.  Alternatively,
they can use the sanitized {\em production traces} we
are making available with this paper~\cite{Trace}.

\myparagraph{Managing cold-start invocations} Using observations from
our characterization, we also propose a practical resource management
policy for reducing the number of cold start executions while
consuming no more resources than the large cloud providers' current
policies.  Specifically, AWS and Azure use a fixed ``keep-alive''
policy that retains the resources in memory for 10 and 20 minutes
after a function execution,
respectively~\cite{Shilkov_AWS_Cold_Starts,Shilkov_Azure_Cold_Starts}.
Though this policy is simple and practical, it disregards the
functions' actual invocation frequency and patterns, and thus behaves
poorly and wastes resources.

In contrast, our policy (1) uses a different keep-alive value for
each user's workload, according to its actual invocation frequency
and pattern; and (2) enables the provider in many cases to pre-warm
a function execution just before its invocation happens (making it a
warm start).  Our policy leverages a small histogram that keeps
track of the recent function inter-invocation times.  For workloads
that exhibit clear invocation patterns, the histogram makes clear
how much keep-alive is beneficial and when the pre-warming should
take place.  For workloads that do not, our policy reverts back to
the fixed keep-alive policy.  As the histogram must be small, for
any workloads that cannot be captured by the histogram but exhibit
predictable invocation patterns, our policy uses time-series
analysis to predict when to pre-warm.

We implement our policy in simulation and for the Apache
OpenWhisk~\cite{openwhisk} FaaS platform, both driven with real
workload traces.  Our simulation results show that the policy
significantly reduces the number of function cold starts, while
spending fewer resources than the fixed keep-alive policy.  Our
experimental results show that the policy can be easily implemented in
real systems with minimal overheads. In fact, we describe our
  recent production implementation in Azure Functions at the end
  of the paper.

\myparagraph{Contributions} In summary, our main contributions are:
\begin{itemize}[wide,labelwidth=!,labelindent=0pt,topsep=0pt,itemsep=-1ex,partopsep=1ex,parsep=1ex]
\item A detailed characterization of the entire production FaaS
  workload at a large cloud provider;
\item A new policy for reducing the number of cold start function
  executions at a low resource provisioning cost;
\item Extensive simulation and experimental results based on real
  traces showing the benefits of the policy;
\item An overview of our implementation in Azure Functions;
\item A large sanitized dataset containing production FaaS traces.
\end{itemize}
\section{Background}
\label{sec:background}

\myparagraph{Abstraction} In FaaS, the user uploads code to the cloud,
and the provider enables a handle (\eg, a URL) for the code to be run.
The choices of which resources to allocate, when to allocate them, and
for how long to retain them, still have to be made, but they are
shifted to the cloud provider.

\myparagraph{Triggers} Functions can be invoked in response to
several event types, called
triggers~\cite{azurefunctionstriggers,awslambdatriggers}.  For
clarity, in this paper we group Azure's many triggers into 7
classes: HTTP, Event, Queue, Timer, Orchestration, Storage, and
others.  Event triggers include Azure Event Hub and Azure Event
Grid, and are used for discrete or serial events, with individual or
batch processing.  Queue-triggered functions respond to message
insertion in a number of message queueing solutions, such as Azure
Service Bus and Kafka.  Timer triggers are similar to cron jobs, and
cause function invocations at pre-determined, regular intervals.  We
grouped all triggers related to Azure Durable
Functions~\cite{azuredurablefunctions} as Orchestration. One can use
these triggers to create native, complex function chaining and
orchestration.  Finally, we grouped database and filesystem triggers
as Storage. These fire in response to changes in the underlying
data, and include Azure Blob Storage and Redis.

\myparagraph{Applications} In Azure Functions, functions
are logically grouped in applications, \ie an application may
encompass multiple functions.  The application concept helps organize
the software and in packaging. \emph{The application, not the
  function, is the unit of scheduling and resource allocation}.

\myparagraph{Cold starts} A \emph{cold start} invocation occurs when a
function is triggered, but its application is not yet loaded in
memory.  When this happens, the platform instantiates a
``worker''~\footnote{In some systems, a worker is a container, but in
  others it can be a VM.}
for the application, loads all the required runtime and libraries, and
calls the function.  This process can take a long time relative to the
function execution~\cite{Wang_2018}.
There are strategies to reduce the time taken by each cold start, such
as keeping pre-allocated VMs or containers, instantiated virtual
network interfaces~\cite{mohan2019agile}, or pre-loaded runtimes that
can be specialized on-demand~\cite{hendrickson2016serverless}. In this
paper, we focus on the complementary and orthogonal goal of reducing
the number of cold starts.

\myparagraph{Concurrency and elasticity} A running instance of an
application can respond to a configurable number of concurrent
invocations of its functions. The number depends on the nature of the
function, and its resource needs.
Cold starts can also happen if there is a spike in the load to an
application, and new instances have to be allocated quickly.  Given
full-server instances and our real FaaS workload, only a tiny
percentage ($<$1\%) of applications would experience this type of cold
start.  For this reason, we do not consider it in this paper.

\myparagraph{Cold start management policy} A key aspect of FaaS is the
trade-off between reducing cold starts by keeping instances warm, and
the resources (\eg, VMs, memory) they need.

Most FaaS providers use a fixed keep-alive policy for all
applications, where application instances are kept loaded in memory
for a fixed amount of time after a function
execution~\cite{Shilkov_AWS_Cold_Starts,Shilkov_Azure_Cold_Starts}.
This is also the case for most open-source implementations (\eg,
OpenWhisk uses a 10-minute period).

This policy is simple to implement and maintain, but does not consider
the wide variety of application behaviors our characterization
unearths.  Thus, it can have many cold starts while wasting resources
for many applications.
Moreover, it is easy to identify by external users, who sometimes
invoke their applications frequently enough (perhaps with dummy
invocations) to keep them warm. This practice amplifies the resource
waste issue.  In this paper, we design a better policy.

\section{FaaS Workloads}
\label{sec:characterization}

We characterize the FaaS workloads seen by
Azure Functions, focusing on characteristics that are intrinsic to the
applications and functions (\eg, their arrival pattern), and not on
the characteristics that relate to the underlying platform (\eg, where
functions are scheduled).  Throughout the characterization, we
highlight interesting observations and their implications for cold
starts and resource management.

\subsection{Data Collection}
\label{sec:collection}

We collected data on all function invocations across Azure's entire
infrastructure between July \nth{15} and July \nth{28}, 2019.  We
collected four related datasets:
\begin{enumerate}[wide,labelwidth=!,labelindent=0pt,topsep=0pt,itemsep=-1ex,partopsep=1ex,parsep=1ex]
\item Invocation counts: per function, in 1-minute bins;
\item Trigger per function;
\item Execution time per function: average, minimum, maximum, and
count of samples, for each 30-second interval, recorded per worker; and 
\item Memory usage per application: sampled every 5 seconds by the
runtime and averaged, for each worker, each minute. Average, minimum,
maximum, and count of samples, for allocated and resident memory.
\end{enumerate}

With this paper, we are releasing a subset of our traces 
at \url{https://github.com/Azure/AzurePublicDataset}.

\myparagraph{Limitations} Given the extreme scale of Azure
Functions, the invocation counts are binned in 1-minute intervals,
\ie our dataset does not allow the precise reconstruction of
inter-arrival times that are smaller than one minute. For this
paper, this granularity is sufficient.

\begin{figure}[t]
\centering
\includegraphics[width=\columnwidth]{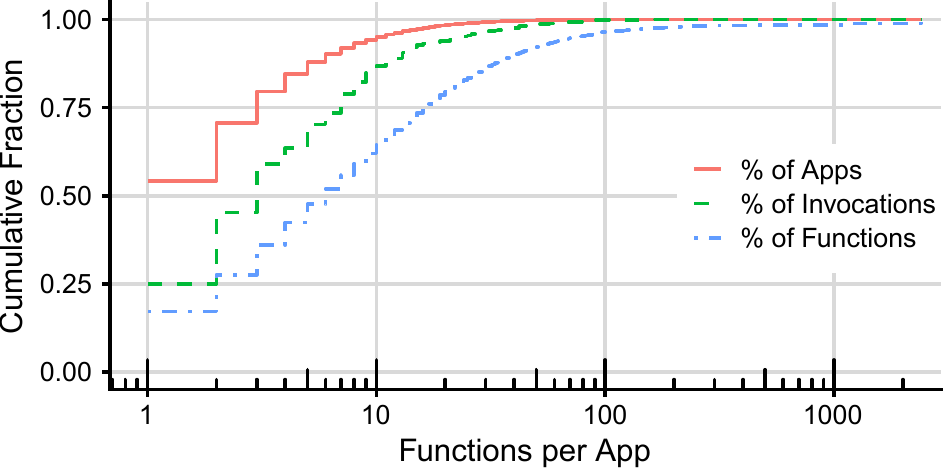}
\vspace{-.25in}
\caption{Distribution of the number of functions per app.}
\label{fig:func-per-app}
\vspace{-.2in}
\end{figure}

For the execution time, we also do not have the complete time
distribution across all invocations. However, from the many samples of
average time, and corresponding counts, we keep a set of weighted
percentiles, where the weight of an entry is the number of samples.
For example, if we see an average time of 100ms over 45 samples, the
resulting percentiles are equivalent to those computed over a
distribution where 100ms are replicated 45 times. The quality of the
approximation to the true distribution depends on the number of
samples in each bin, the smaller the better. We similarly obtain
weighted percentiles for memory usage.

For confidentiality reasons, we cannot disclose some absolute numbers,
such as total number of functions and invocations. Nevertheless, our
characterization is useful for understanding a full FaaS workload, and
for researchers and practitioners to generate realistic FaaS
workloads.

\begin{figure}[t]
\centering
\includegraphics[width=2in, trim=1mm 1mm 1mm 1mm, clip = true]{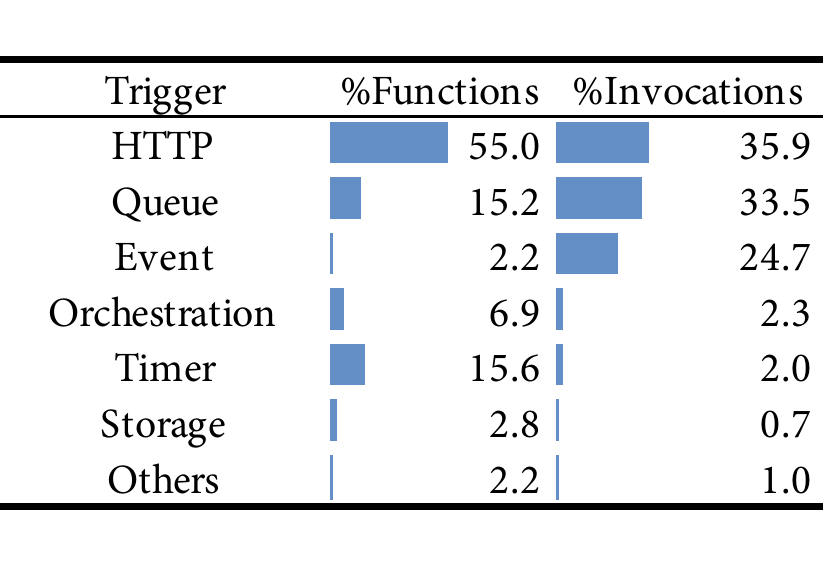}
\vspace{-.2in}
\caption{Functions and invocations per trigger type.}
\label{fig:triggers-per-function}
\vspace{-.2in}
\end{figure}

\subsection{Functions, Applications, and Triggers}

\myparagraph{Functions and applications} Figure~\ref{fig:func-per-app}
shows the CDF of the number of functions per application (top curve).
We observe that 54\% of the applications only have one function, and
95\% of the applications have at most 10 functions. About 0.04\% of
the applications have more than 100 functions.

The other two curves show the fraction of invocations, and functions,
corresponding to applications with up to a certain number of
functions.  For example, we see that 50\% of the invocations come
from applications with at most 3 functions, and 50\% of the functions
are part of applications with at most 6 functions. Though we found a
weak positive correlation between the number of functions in an
application and the median number of invocations of those
applications, the number of functions in an application is not a
useful signal in resource management.

We took a closer look at the 10 applications with the most functions.
Only 4 had more than 1k functions: these, and 3 others, had a
pattern of auto-generated function names triggered by timers or
HTTP, which suggests that they were being used for large automated
testing. Of the remaining 3 applications, two were
using Azure Durable Functions for orchestrating multiple functions,
and one seems to be an API application, where each function
corresponds to one route in a large Web or REST application. 
We plan to do a broader and more comprehensive
study of application patterns in future work.

\myparagraph{Triggers and applications}
Figure~\ref{fig:triggers-per-function} shows the fraction of all
functions, and all invocations, per type of trigger.  HTTP is the most
popular in both dimensions. Event triggers correspond to only 2.2\% of
the functions, but to 24.7\% of the invocations, due to their
automated, and very high, invocation rates. Queue triggers also have
proportionally more invocations than functions (33.5\% vs 15.2\%). The
opposite happens with timer triggers. There are many functions
triggered by timers (15.6\%), but they correspond to only 2\% of the
invocations, due to the relatively low rate they fire in: 95\% of the
timer-triggered functions in our dataset were triggered at most once per
minute, on average.

\begin{figure}[t]
\centering
\subfloat[Apps with $\ge 1$ of each trigger.]
	{\raisebox{1cm}{\includegraphics[width=4cm]{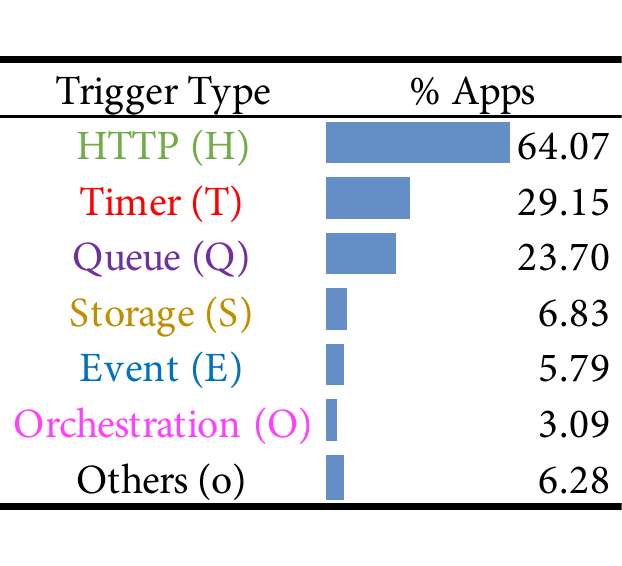}}}\quad
\subfloat[Popular trigger combinations.]
	{\includegraphics[width=4cm]{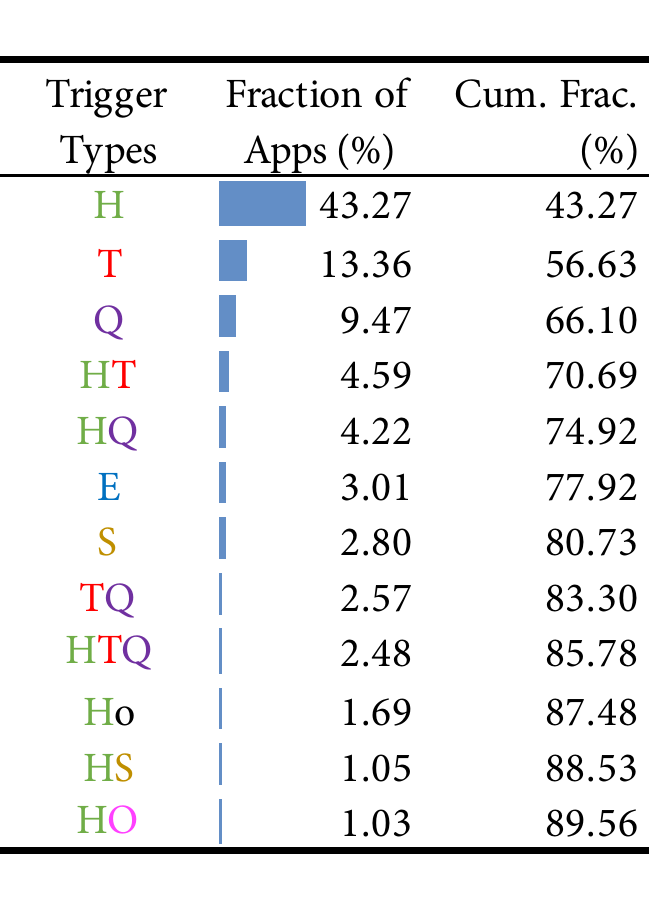}}
\vspace{-.1in}
\caption{Trigger types in applications.}
\label{fig:triggers-groups-per-app}
\vspace{-.15in}
\end{figure}

Figure~\ref{fig:triggers-groups-per-app} shows how applications
combine functions with different trigger types.  In
Figure~\ref{fig:triggers-groups-per-app}(a), we show the applications
with at least one trigger of the given type. We find that 64\% of the
applications have at least one HTTP trigger, and 29\% of the
applications have at least one timer trigger. As applications can have
multiple triggers, the fractions sum to more than 100\%.  In
Figure~\ref{fig:triggers-groups-per-app}(b), we partition the
applications by their combinations of triggers. 43\% of the
applications have \emph{only} HTTP triggers, and 13\% of the apps have
\emph{only} timer triggers. Combining the two tables, we find that
15.8\% of the applications have timers and at least one other trigger
type. For predicting invocations, as we discuss later, while timers
are very predictable, 86\% of the applications have either no timers
or timers combined with other triggers.

\subsection{Invocation Patterns}

\begin{figure}[t]
\centering
\includegraphics[width=\columnwidth]{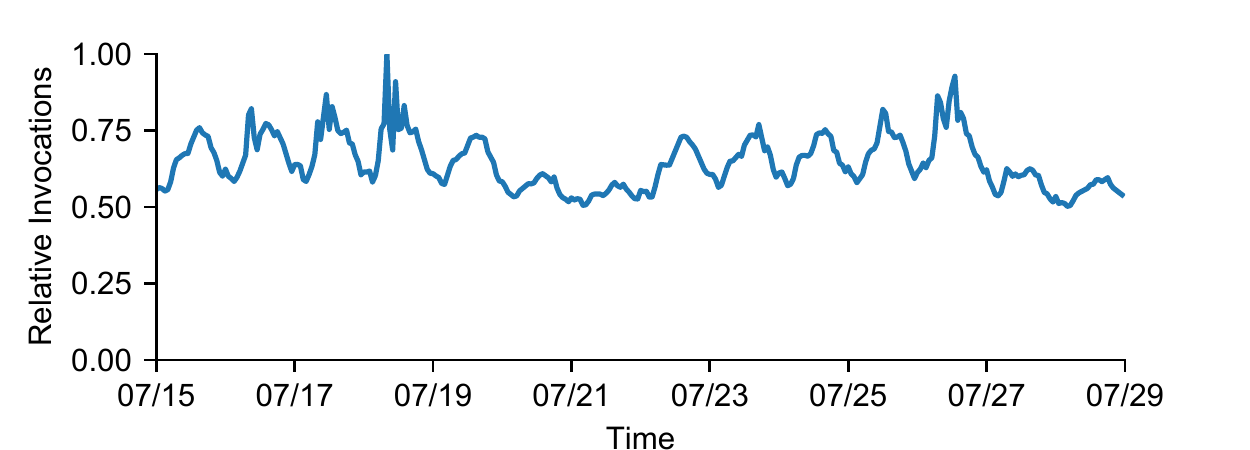}
\vspace{-.25in}
\caption{Invocations per hour, normalized to the peak.}
\label{fig:invocations-per-hour}
\vspace{-.2in}
\end{figure}

We now look at dynamic function and application invocations.
Figure~\ref{fig:invocations-per-hour} shows the volume of invocations
per hour, across the entire platform, relative to the peak hourly load
on July \nth{18}. There are clear diurnal and weekly patterns (July
\nth{20}, \nth{21}, \nth{27}, and \nth{28} are weekend days), and a
constant baseline of roughly 50\% of the invocations that does not
show variation.  Though we did not investigate this specifically,
there can be several causes, \eg a combination of human and
machine-generated traffic, plain high-volume applications, or the
overlapping of callers in different time zones.

\begin{figure}
\subfloat[CDF of daily invocations per function and application, and
  the corresponding \emph{average} interval between invocations.
  Shaded regions show applications invoked on average at most once per
  hour (green, 45\% of apps) and at most once per minute (yellow, 81\%
  of
  apps).]{\includegraphics[width=\columnwidth]{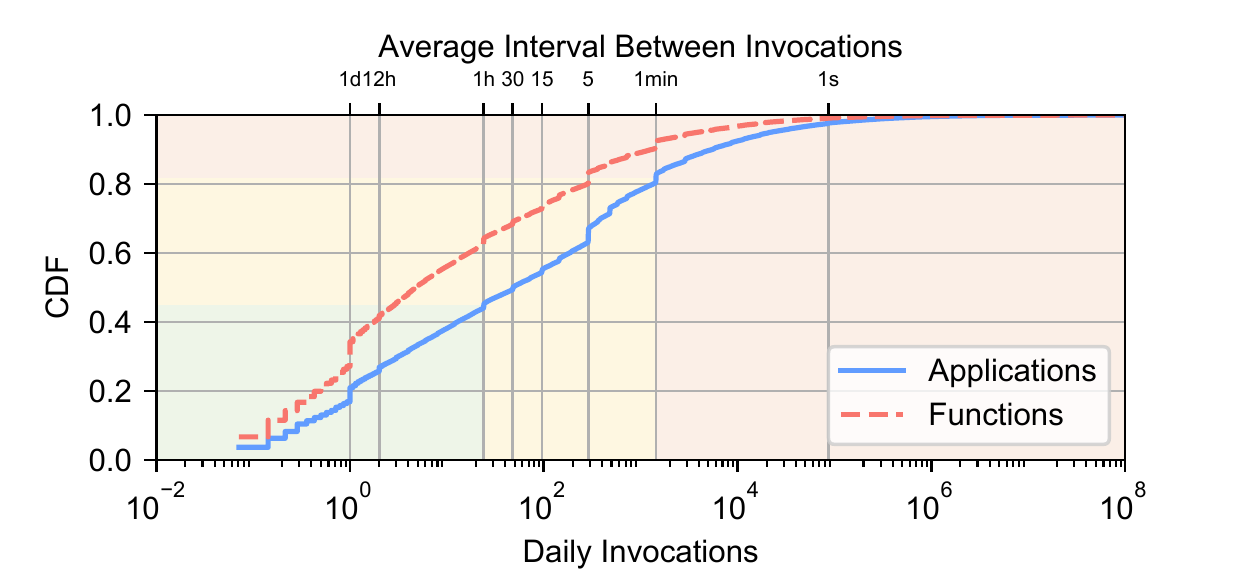}}
\\ \subfloat[Fraction of total function invocations by the fraction of
  the most popular functions and applications. Same colors as in
  Figure~\ref{fig:invocations-per-app}(a).]
  {\includegraphics[width=\columnwidth]{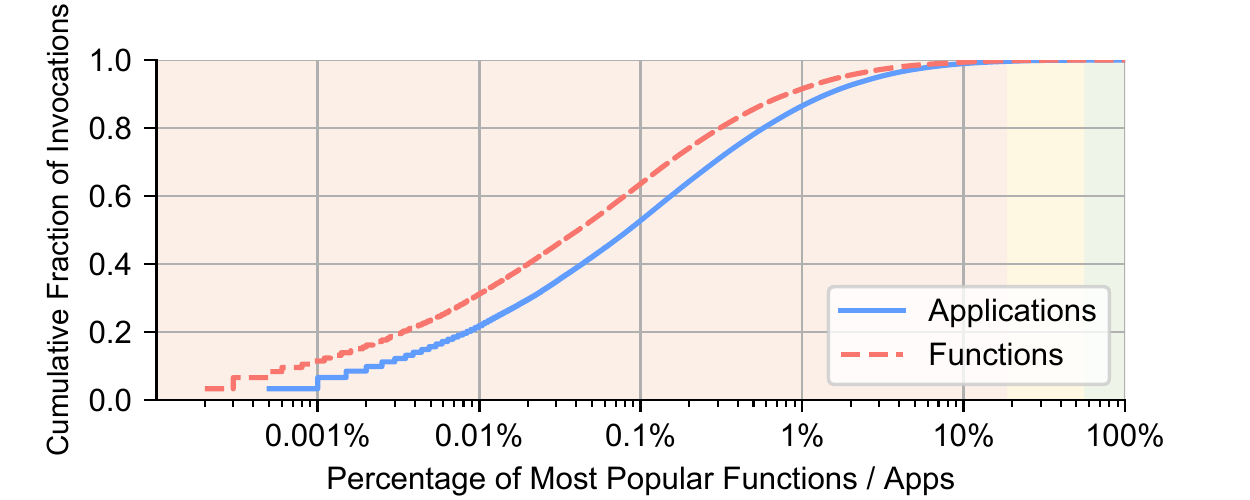}}
\vspace{-.1in}
\caption{Invocations per application and per function for a
representative sample of the dataset.}
\label{fig:invocations-per-app}
\vspace{-.175in}
\end{figure}

Figure~\ref{fig:invocations-per-app}(a) shows the CDF of the average
number of invocations per day, for a representative sample of 
both functions and applications.
The invocations for an application are the sum
over all 
its functions. First, we see that the number of invocations per
day varies by over 8 orders of magnitude for functions and
applications, making the resources the provider has to dedicate to
each application also highly variable.

The second observation with strong implications for resource
allocation is that the vast majority of applications and functions are
invoked, on average, very infrequently. The green- and yellow-shaded
areas in the graph show, respectively, that 45\% of the applications
are invoked once per hour or less on average, and 81\% of the
applications are invoked once per minute or less on average. This
suggests that the cost of keeping these applications warm, relative to
their total execution (billable) time, can be prohibitively high.

Figure~\ref{fig:invocations-per-app}(b) shows the other side of the
workload skewness, by looking at the cumulative fraction of
invocations due to the most popular functions and applications
in the sample. The
shaded areas correspond to the same applications as in
Figure~\ref{fig:invocations-per-app}(a).  The applications in the
orange-shaded area are the 18.6\% most popular, those invoked on
average at least once per minute. They represent 99.6\% of
all function invocations.

The invocation rates provide information on the average inter-arrival
time (IAT) of function and application invocations, but not on the
distribution of these IATs. If the next invocation time of a function
can be predicted, the platform can avoid cold starts by pre-warming
the application right before it is to be invoked, and save resources
by shutting it down right after execution.

\myparagraph{Inter-arrival time variability} To gain insight into the
IAT distributions of applications, we look at the coefficient of
variation (CV) of each application. The CV (standard deviation divided
by the mean) provides a measure of the variability in the IATs. We
would expect timer-triggered functions to have periodic arrivals, with
a CV of 0. Human-generated invocations should approximately follow a
Poisson arrival process, with an exponential (memoryless) distribution
of IATs~\cite{gallager13stochastic}. These would ideally yield a CV of
1.  CVs greater than 1 suggest significant variability.

\begin{figure}[t]
\centering
\includegraphics[width=\columnwidth]{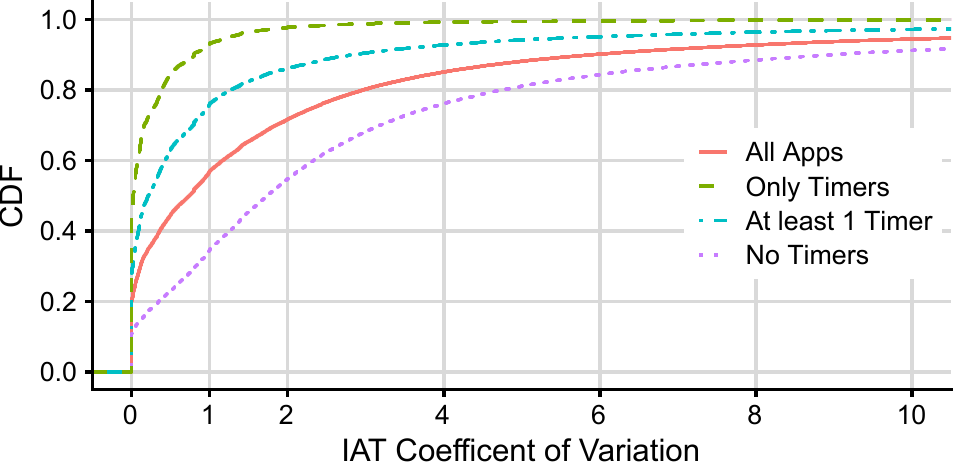}
\vspace{-.25in}
\caption{CV of the IATs for subsets of applications.}
\label{fig:iat-variation}
\vspace{-.2in}
\end{figure}

Figure~\ref{fig:iat-variation} shows the distribution of the CV across
all applications, as well as for subsets of applications with and
without timers.  It shows that the real IAT distributions are more
complex than the simply periodic or memoryless ones. For example, only
$\sim$50\% of the applications with only timer-triggered functions
have a CV of 0. Multiple timers with different periods and/or phases
will increase the CV. For applications with at least one timer, this
fraction is less than 30\%, and across all applications the fraction
is $\sim$20\%. Interestingly, $\sim$10\% of applications with no
timers have CV close to 0, which means they are quite periodic, and
should be predictable. This could be due to, for example, external
callers (e.g., sensors or IoT devices) that operate periodically. On
the other hand, only a small fraction of applications has a CV close
to 1, meaning that simple Poisson arrivals are not the norm.  These
results show that there is a significant fraction of applications that
should have fairly predictable IATs, even if they do not have timer
triggers.  At the same time, these numbers suggest that for many
applications predicting IATs is not trivial.

\subsection{Function Execution Times}

Another aspect of the workload is the function execution time, \ie the
time functions take to execute \emph{after they are ready to run}. In
other words, these numbers do not include the cold start times. Cold
start times depend on the infrastructure to a large extent, and have
been characterized in other studies~\cite{Wang_2018}.

\begin{figure}[t]
\centering
\includegraphics[width=\columnwidth]{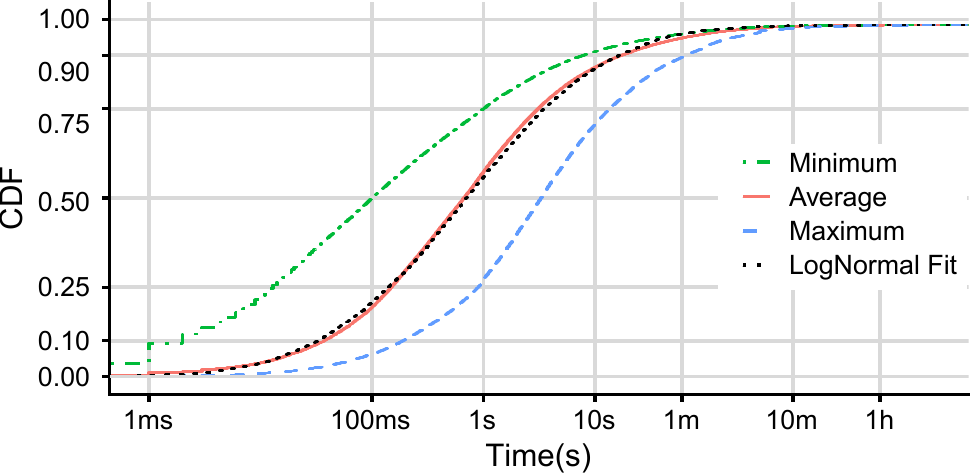}
\vspace{-.3in}
\caption{Distribution of function execution times. Min, avg, and max
  are separate CDFs, and use independent sorting.}
\label{fig:function-duration}
\vspace{-.2in}
\end{figure}

Figure~\ref{fig:function-duration} shows the distribution of average,
minimum, and maximum execution times of all function executions on
July \nth{15}, 2019. The distributions for other days are similar. The
graph also shows a very good log-normal fit (via MLE) to the
distribution of the averages, with log mean -0.38 and $\sigma$
2.36. We observe that 50\% of the functions execute for less than 1s
on average, and 50\% of the functions have maximum execution time
shorter than $\sim$3s; 90\% of the functions take at most 60s, and
96\% of functions take less than 60s on average.

The main implication is that the function execution times are at the
same order of magnitude as the cold start times reported for major
providers~\cite{Wang_2018}. \emph{This makes avoiding and/or
  optimizing cold starts extremely important for the overall
  performance of a FaaS offering.}

Another interesting observation is that, overall, functions in this
FaaS workload are very short compared to other cloud workloads.  For
example, data from Azure~\cite{resourcecentral} shows that 63\% of all
VM allocations last longer than 15 minutes, and only less than 8\% of
the VMs last less 5 minutes or less.  This implies that FaaS imposes
much more stringent requirements on the provider to stand-up resources
quickly.

\myparagraph{Idle times} As we discuss in Section~\ref{sec:policy}, an
important aspect of the workload for managing cold starts is idle time
(IT), defined as the time between the end of a function's execution
and its next invocation.  IT relates to IAT and execution time.  For
most applications, the average execution time is at least 2 orders of
magnitude smaller than the average IAT. We verified for the
applications in the yellow region in
Figure~\ref{fig:invocations-per-app}(a) -- 81\% of the applications
invoked at most once per minute on average -- that indeed the IT and
IAT distributions are extremely similar.

\myparagraph{Potential correlations} Different triggers had average
function execution times differing by about 10$\times$, between
200ms and 2s at the median, but all with the same shape for the
distributions. One outlier was a class of orchestration functions
with median average execution times of $\sim$30ms, as they simply
dispatch and coordinate other functions.

\subsection{Memory Usage}  

\begin{figure}[t]
\centering
\includegraphics[width=\columnwidth]{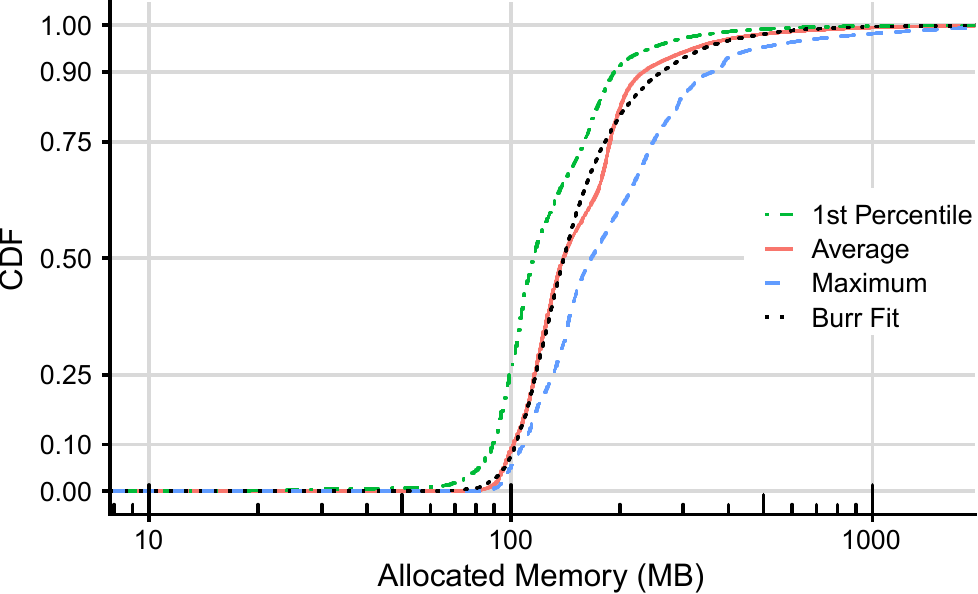}
\vspace{-.3in}
\caption{Distribution of allocated memory per application.}
\label{fig:app-memory}
\vspace{-.2in}
\end{figure}

We finally look at the memory demands of applications. Recall that
the application is the unit of memory allocation in the platform we
study.  Figure~\ref{fig:app-memory} shows the memory demand
distribution, across all applications running on July \nth{15},
2019. We present three curves drawn from the memory data: \nth{1}
percentile, average, and maximum allocated memory for the
application.  We also plot a reasonably good Burr distribution fit
(with parameters $c=11.652$, $k=0.221$, and $\lambda=107.083$) for
the average.  Allocated memory is the amount of virtual memory
reserved for the application, and may not necessarily be all
resident in physical memory.  Here, we use the \nth{1} percentile
because there was a problem with the measurement of the minimum,
which made that data not usable.  Despite the short duration of each
function execution, applications tend to remain resident for longer.
The distributions for other days in the dataset are very similar.

Looking at the distribution of the maximum allocated memory, 90\% of
the applications never consume more than 400MB, and 50\% of the
applications allocate at most 170MB. Overall, there is a 4$\times$
variation in the first 90\% of applications, meaning that memory is an
important factor in warmup, allocation, and keep-alive decisions for
FaaS.

\myparagraph{Potential correlations} We found no strong correlation
between invocation frequency and memory allocation or between memory
allocation and function execution times.

\subsection{Main Takeaways}

From the point of view of cold starts and resource allocation, we now
reiterate our three main observations.  First, the vast majority of
functions execute on the order of a few seconds -- 75\% of them have a
maximum execution time of 10 seconds -- so execution times are on the
same order as the time it takes to start functions cold.  Thus, it is
critical to reduce the number of cold starts or make cold starts
substantially faster.  Eliminating a cold start is the same as making
it infinitely fast.

Second, the vast majority of applications are invoked infrequently --
81\% of them average at most one invocation per minute.  At the same
time, less than 20\% of the applications are responsible for 99.6\% of
all invocations.  Thus, it is expensive, in terms of memory footprint,
to keep the applications that receive infrequent invocations resident
at all times.

Third, many applications show wide variability in their IATs -- 40\%
of them have a CV of their IATs higher than 1 -- so the task of
predicting the next invocation can be challenging, especially for
applications that are invoked infrequently.

\section{Managing Cold Starts in FaaS}
\label{sec:policy}

We use insights from our characterization to design an adaptive
resource management policy, called {\em hybrid histogram policy}.
The goal is to reduce the number of cold start invocations with
minimum resource waste.  We refer to a \emph{policy} as a set of rules
that govern two parameters {\em for each application}:

\noindent\emph{--- Pre-warming window.}  The time the policy waits,
since the last execution, before it loads the application image
expecting the next invocation.  A pre-warming window = 0 means that
the policy does not unload the application after one of its functions
executes.  Aggressive pre-warming (a large window) reduces resource
usage but may also cause cold starts, in case the next invocation
occurs sooner than expected.

\noindent\emph{--- Keep-alive window.}  The time during which an
application's image is kept alive after (1) it has been loaded to
memory (pre-warming window $\geq$ 0) or (2) a function execution
(pre-warming window = 0).  (Note that our definition for this window
differs from the keep-alive parameter in fixed keep-alive policies,
which is the same for all applications and only starts at the end of
function executions.)  Longer windows have the potential to reduce
cold starts by increasing the chances of an invocation falling into
this window.  However, this may also waste resources, \ie leave them
idle, in case the next invocation does not happen soon after loading.

A \emph{no-unloading policy} would keep every application image loaded
in memory all the time (\ie, infinite keep-alive window and
pre-warming window = 0).  This policy would get no cold starts but
would be too expensive to operate.

\subsection{Design Challenges}

Designing a \emph{practical policy} poses several challenges:
\begin{enumerate}[wide,labelwidth=!,labelindent=0pt,topsep=0pt,itemsep=-1ex,partopsep=1ex,parsep=1ex]
\item {\bf Hard-to-predict invocations.} As
  Figure~\ref{fig:triggers-groups-per-app} shows, many applications
  are triggered by timers.  A timer-aware policy could leverage this
  information to pre-warm applications right before the next
  invocation.  However, predicting the next invocation is challenging
  for other triggers.

\item {\bf Heterogeneous applications.} As
  Figure~\ref{fig:invocations-per-app} shows, the invocation frequency
  and pattern vary substantially across applications.  A
  one-size-fits-all fixed policy is certain to be a poor choice for
  many applications.  A better policy should adapt to each application
  dynamically.

\item {\bf Applications with infrequent invocations.} Some
applications are invoked very infrequently, so an adaptive policy
would take some time to learn their invocation patterns.  The same
applies to applications that it sees for the first time.

\item {\bf Tracking overhead.} Adapting the policy to each
application means tracking each application individually.  For this
reason, the cost to track the information for each application should
be small.  For example, we need to consider the size of the data
structures that will keep this state.

\item {\bf Execution overhead.} Since function executions can be very
  short (\ie, more than 50\% of executions take less than 1 second),
  running the policy and updating its state need to be fast.  This is
  especially critical considering providers charge users only during
  their function execution times (\eg, based on CPU, memory).  For
  instance, we cannot take 100 ms to update a policy for each 10
  ms-long execution.  Due to these overheads, expensive prediction
  techniques, such as time-series analysis, cannot be used for all
  applications.
\end{enumerate}

\subsection{Hybrid Histogram Policy}
\label{subsec:hybrid_hist_policy}

\myparagraph{Overview} Our hybrid histogram policy addresses all the
above challenges.  To address challenges \#1 and \#2, our policy
adjusts to the invocation frequencies and patterns of each individual
application.  It identifies the application's invocation pattern,
removes/unloads the application right after each function execution
ends, reloads/pre-warms the application right before a potential next
invocation (after a ``pre-warming window'' elapses), and keeps it
alive for a period (until a ``keep-alive window'' elapses).  The
pre-warming window starts
after each function execution, and the
keep-alive window starts after each pre-warming.  If the pre-warming
window is 0, we do not unload the application after an execution,
and the end of the execution still starts
a new keep-alive window.  We
explain how exactly we compute the length of these windows below.

\begin{figure}[!t]
    \centering
    \includegraphics[width=0.8\linewidth]{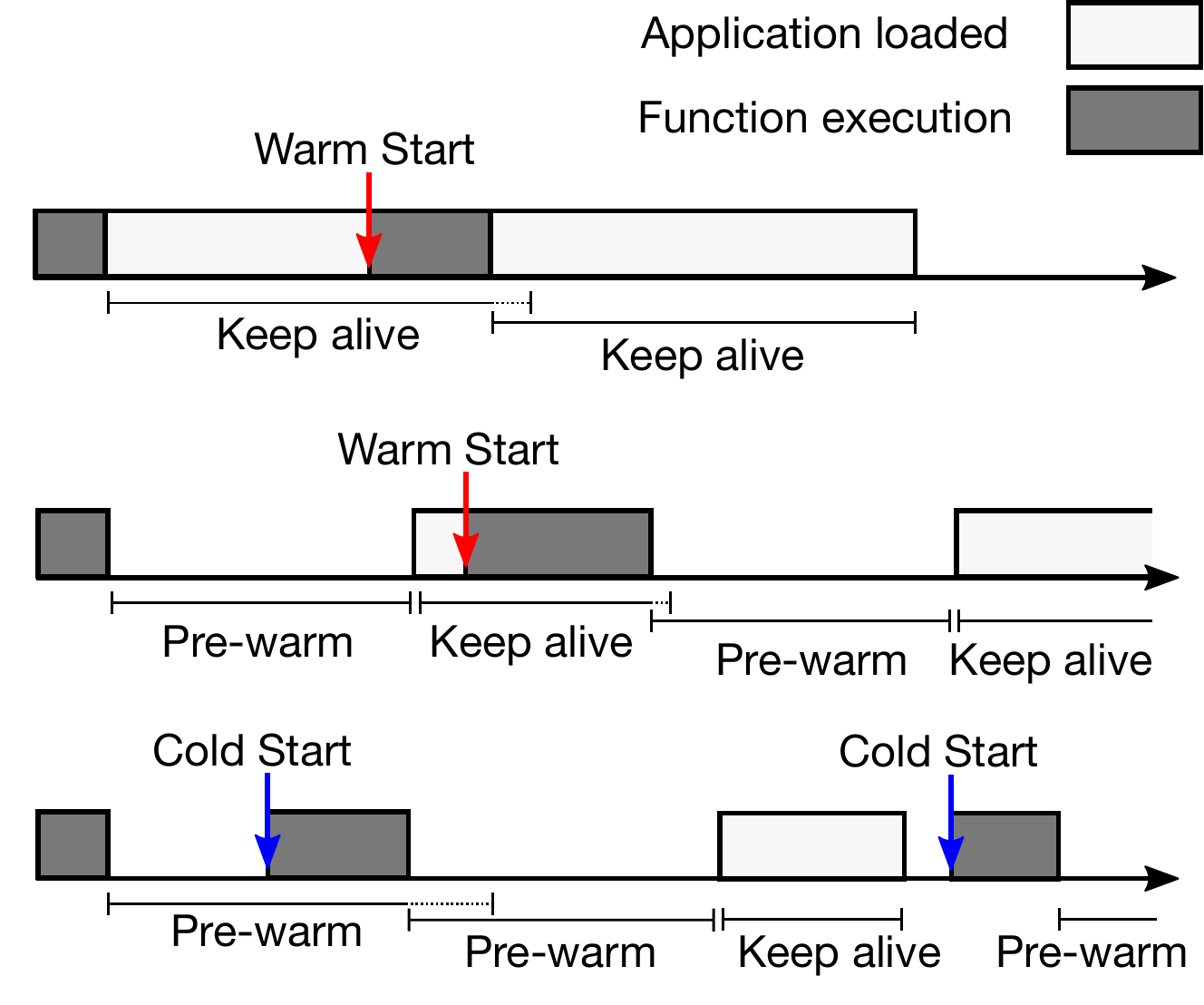}
    \vspace{-.1in}
    \caption{Timelines showing a warm start with keep alives and no
pre-warming (top); a warm start following a pre-warm (middle); and
two cold starts, before a pre-warm, and after a keep alive (bottom).}
    \label{fig:windows}
    \vspace{-.2in}
\end{figure}

Figure~\ref{fig:windows} shows the pre-warming and keep-alive windows in
three scenarios.  In the top scenario, the pre-warming window is 0,
and an invocation that happens before the keep-alive window ends is
a warm start. The end of the execution starts a new
keep-alive window. In the middle, the next invocation is a warm
start, as the application is re-loaded after a pre-warming
window. The end of the execution starts a new pre-warming
window. In the bottom scenario, there are two cold starts: the first 
resulting from an invocation arriving before the pre-warming window
elapsed, and the second from an invocation arriving after the
keep-alive period
elapsed.

\begin{figure}[!t]
    \centering
    \includegraphics[width=\linewidth]{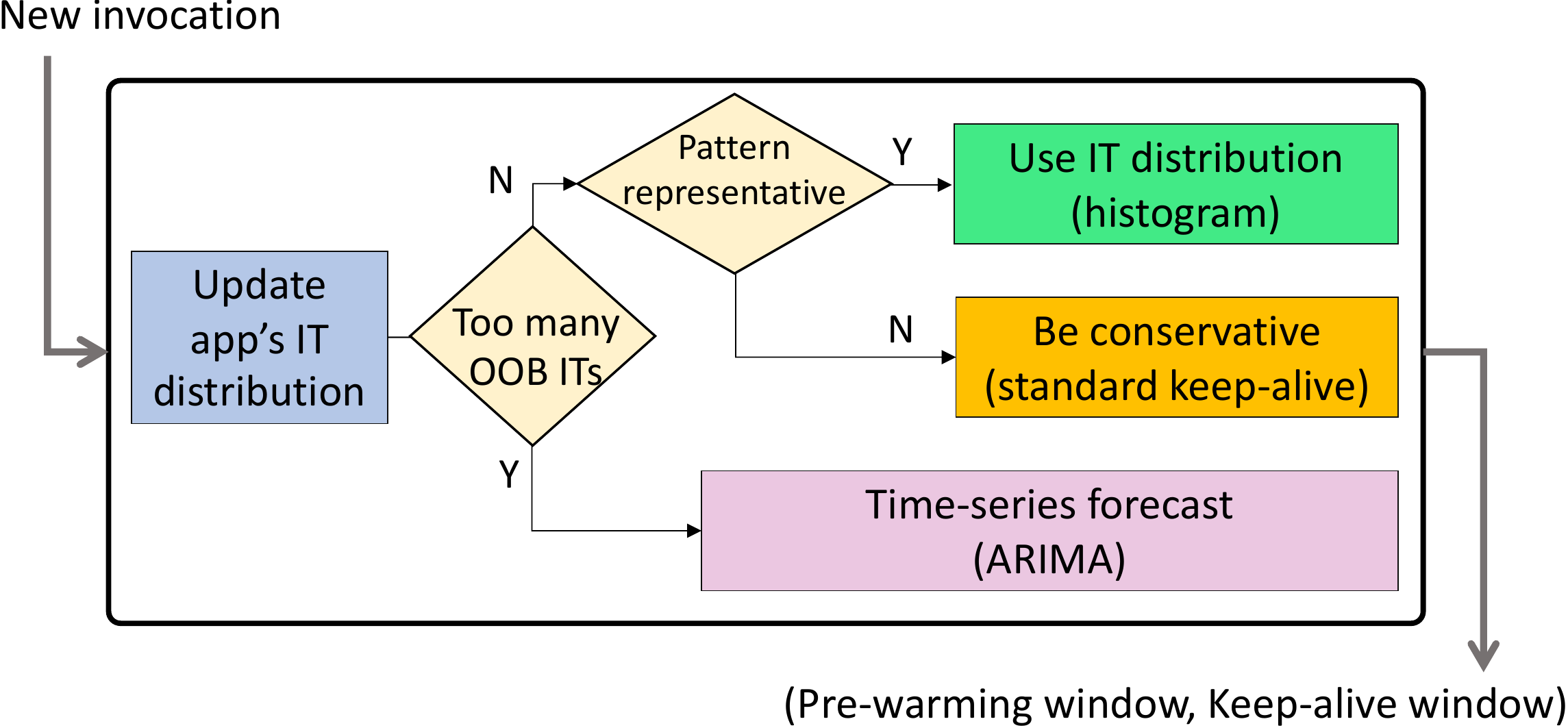}
    \vspace{-.25in}
    \caption{Overview of the hybrid histogram policy.}
    \label{fig:histogram_flow_chart}
    \vspace{-.05in}
\end{figure}

The policy comprises three main components: (1) a range-limited
histogram for capturing each application's ``idle'' times (ITs);
(2) a
standard keep-alive approach for when the histogram is not
representative, \ie there are too few ITs or the IT behavior is
changing (again, note that this differs from a fixed keep-alive
policy); and (3) a time-series forecast component for when the
histogram does not capture most ITs.
Figure~\ref{fig:histogram_flow_chart} overviews our policy and its
components. Ultimately, the policy defines the pre-warming and
keep-alive windows for each application.  Next, we describe each
component in turn.

\myparagraph{Range-limited histogram} To address challenges \#4 and
\#5, the centerpiece of our policy is a compact histogram data
structure that tracks the IT distribution for each application.  Each
entry/bin of the histogram counts the number of ITs of the
corresponding length that have occurred.  We use 1-minute bins, which
strikes a good balance between metadata size and the resolution needed
for policy actions.  Keep-alive time scales are in orders of minutes
for FaaS platforms.  We use the same scale for pre-warming.  In
addition, the histogram tracks ITs of up to a configurable duration
(\eg, 4 hours).  Any ITs longer than this are considered ``out of
bounds'' (OOBs).

\begin{figure}[!t]
    \centering
    \includegraphics[width=0.9\linewidth]{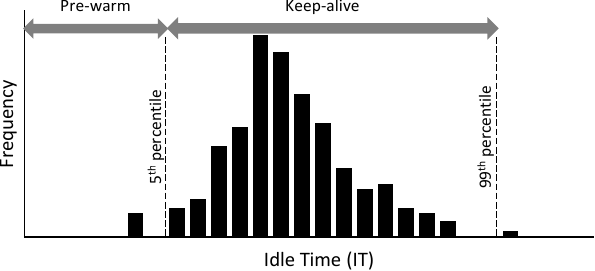}
    \vspace{-.1in}
    \caption{Example application idle time (IT) distribution used to
      select pre-warming times and keep-alive windows.}
    \label{fig:histogram_intro}
    \vspace{-.2in}
\end{figure}

Given the ITs that are within bounds, our policy identifies the head
and tail of the IT distribution.  We use the head to select the
pre-warming window for the application, and the tail to select the
keep-alive window.
To exclude outliers, we set the head and tail by default to the
\nth{5}- and \nth{99}-percentiles of the IT distribution.  (When one
of these percentiles falls within a bin, we ``round'' it to the next
lower value for the head or the next higher value for the tail.)
These two configurable thresholds strike a balance between managing
cold starts and resource costs.  Figure~\ref{fig:histogram_intro}
shows the histogram for a sample application, and the head and tail
markers.  To give the policy a little room for error, our
implementation uses a 10\% ``margin'' by default, \ie it reduces the
pre-warming window by 10\% and increases the keep-alive window by
10\%.

\begin{figure}[!t]
    \centering
    \includegraphics[width=1\linewidth]{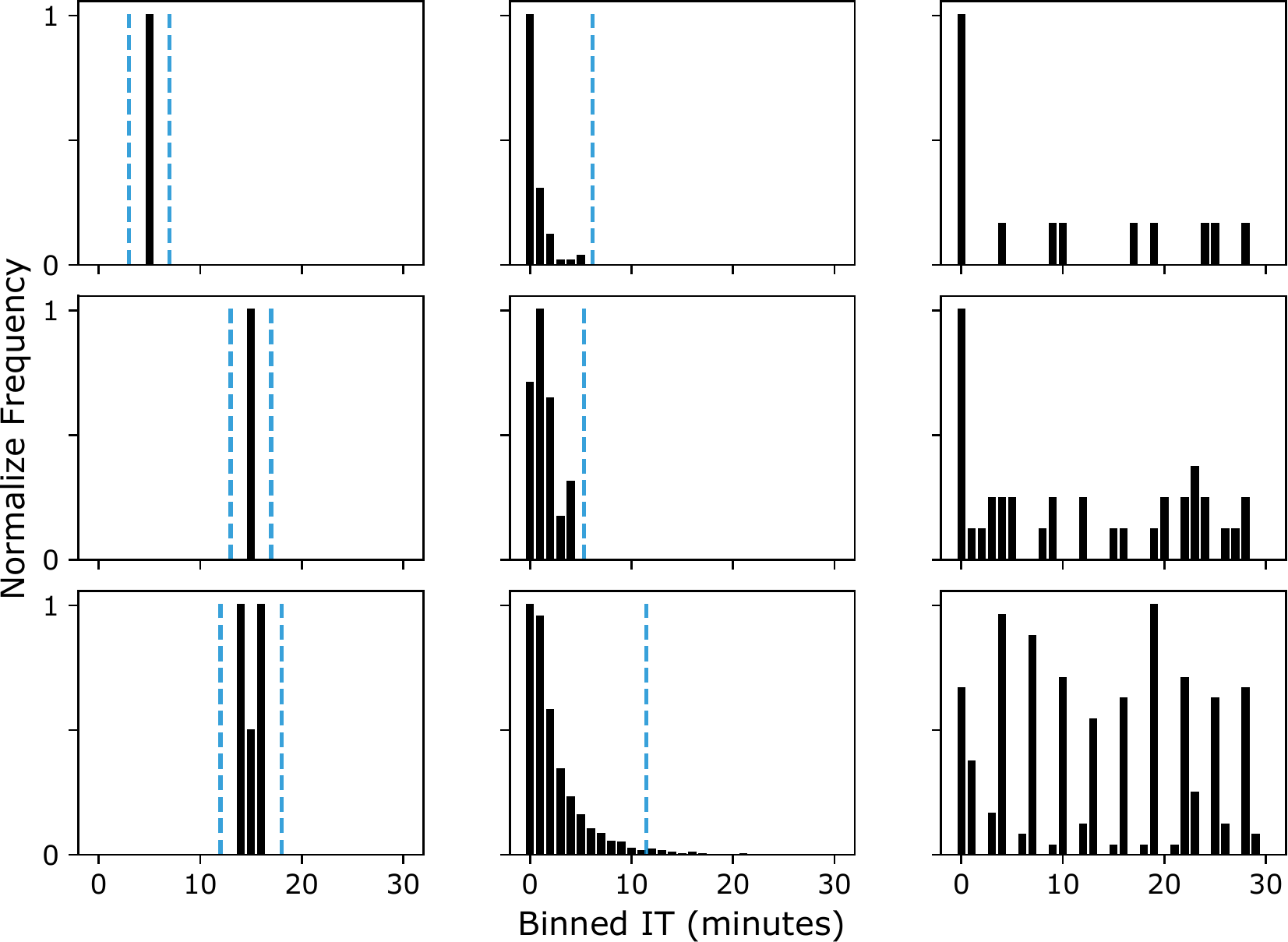}
    \vspace{-.25in}
    \caption{Nine normalized IT distributions from real FaaS workloads
      over a week.}
    \label{fig:sample_histograms}
    \vspace{-.2in}
\end{figure}

Figure~\ref{fig:sample_histograms} shows nine real IT distributions
over a week.  The three histograms in the left column show cases where
both head and tail cutoffs are easy to identify.  These distributions
produce the ideal situation: long pre-warm windows and short
keep-alive windows.  The center cases show no head cutoff as the head
marker rounded down to 0.  In these cases, the pre-warming window is 0
and the policy does not kill the application after a function
execution.

\myparagraph{Standard keep-alive when the pattern is uncertain} The
histogram might not be representative of an application's behavior
when (1) it has not observed enough ITs for the application, or (2)
when the application is transitioning to a different IT regime (\eg,
change from a consistent pattern to an entirely new one).  When the
histogram is not representative, we revert to a standard keep-alive
approach: pre-warming window = 0 and keep-alive window = range of the
histogram (\eg, 4 hours).  This conservative choice of keep-alive
window seeks to minimize the number of cold starts while the histogram
is learning a new pattern.  Our policy reverts back to using the
histogram when it becomes representative (again).

We decide whether a histogram is representative by computing the CV of
its bin counts.  A histogram that has a single bin with a high count
and all others 0 would have a high CV, whereas a histogram where all
bins have the same value would have CV = 0.  The histogram is most
effective in the former case, where there is a large concentration of
ITs (left and center of Figure~\ref{fig:sample_histograms}).  It is
not as effective when ITs are spread widely (right of
Figure~\ref{fig:sample_histograms}).  Thus, if the CV is lower than a
threshold, we use the standard keep-alive approach.  To track the CV
efficiently, we use Welford's online algorithm~\cite{Welford_1962}.

\myparagraph{Time-series analysis when histogram is not large enough}
A compact histogram cannot represent ITs larger than its range.  Thus,
applications with very infrequent invocations (challenge \#3) may
exhibit many out-of-bounds ITs.  For these applications, our policy
uses time-series analysis to predict the next IT.  Specifically, we
use ARIMA modeling~\cite{box1970distribution}.

With an IT prediction, our policy sets the pre-warm window to elapse
just before the next invocation and a short keep-alive window. In more
detail, we used the \texttt{auto\_arima} implementation from the
\texttt{pmdarima} package~\cite{pmdarima_1_2_1_2019}, which
automatically searches for the ARIMA parameters ($p,d,q$) that produce
the best fit.  As applications using ARIMA are invoked very
infrequently, we update the model for each of them after every
invocation.  To give the prediction some room more inaccuracy, we
include a (configurable) margin of 15\%. For example, if the predicted
IT is 5 hours, we set the pre-warming window to 4.25 hours (5 hours
minus 15\%) and the keep-alive window to 1.5 hours (15\% of 5 hours in
each side of the IT prediction).

\myparagraph{Justification} Like other FaaS cold start policies, our
policy eagerly frees up memory when it is not needed.  An alternative
would have been to leverage standard (lazy) caching policies, which
free up cache space only on-demand.  Section~\ref{sec:related}
explains the differences between these types of policies that justify
our approach.  Our policy uses a standard keep-alive with a long
window, when it does not have accurate IT data about the application,
to conservatively prevent cold starts.  A shorter window would lower
cost but would incur more cold starts.  We prefer our approach because
it often quickly reduces memory usage greatly, after the histogram
becomes active for the application.  Instead of using a histogram, we
could attempt to predict the next invocation or idle time using
time-series analysis or other prediction models.  We experimented with
some models, including ARIMA, but found them to be inaccurate or
excessively expensive for the bulk of invocations.  The histogram is
accurate, compact, and fast to update.  So, we rely on ARIMA only for
the applications that cannot be represented with a compact histogram.
Producing an ARIMA model is expensive, but can be off the critical
path.  Moreover, these applications involve only a small percentage of
invocations, so computation needs are kept small.  Nevertheless, we
can easily replace ARIMA with another model.

\subsection{Implementation in Apache OpenWhisk}
\label{sec:implementation}

We implement our policy in Apache OpenWhisk~\cite{openwhisk}, which is
the open-source FaaS platform that powers IBM's Cloud
Functions~\cite{ibmfunctions}. It is written in Scala.

\begin{figure}[!t]
    \centering
    \includegraphics[width=0.9\linewidth]{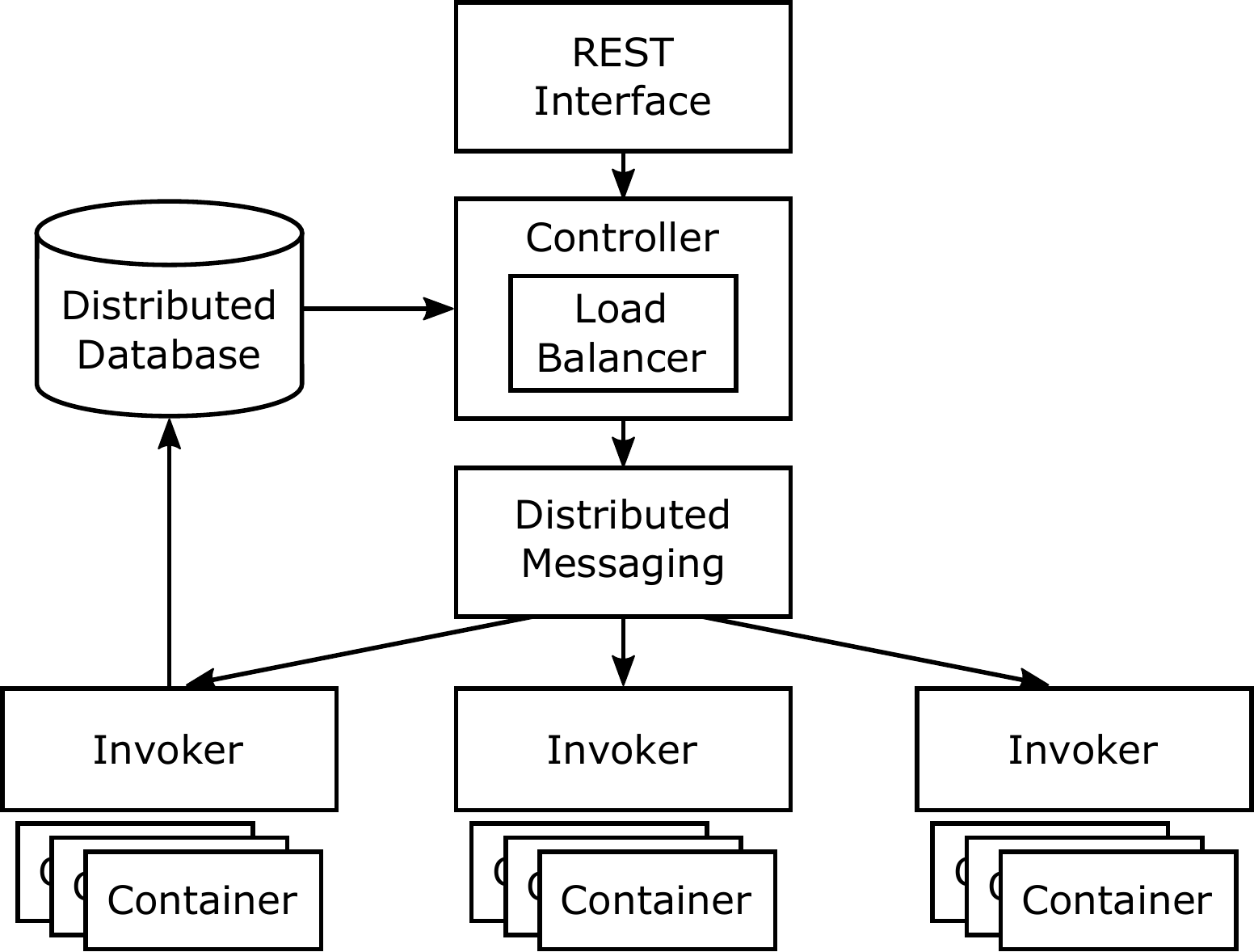}
    \vspace{-.05in}
    \caption{OpenWhisk architecture.}
    \label{fig:openwhisk-architecture}
    \vspace{-.2in}
\end{figure}

\myparagraph{OpenWhisk architecture}
Figure~\ref{fig:openwhisk-architecture} shows the architecture of
OpenWhisk~\cite{openwhisk-architecture}.  It exposes a REST interface
(implemented using Nginx) for users to interact with the FaaS
platform.  A user can create new functions (\emph{actions} in
OpenWhisk terminology), submit new invocations (\emph{activations} in
OpenWhisk terminology), or query their status.  Here, we focus on
function invocation and container management.  Invocation requests are
forwarded to the Controller component, who decides which Invoker
should execute each function instance.  This logic is implemented in
the Load Balancer, which considers the health and available capacity
of the Invokers, as well as the history of prior executions.  The
Controller sends the function invocation request to the selected
Invoker via the distributed messaging component (implemented using
Kafka).  The Invoker receives the invocation request, starts the
function in a Docker container, and manages its runtime (including
when to stop the container).
By default, each Invoker implements a fixed 10-minute keep-alive
policy, and informs the Controller when it unloads a container.

\myparagraph{Implementing our policy} We modify the following
OpenWhisk components to implement the hybrid policy:
\begin{enumerate}[wide,labelwidth=!,labelindent=0pt,topsep=0pt,itemsep=-1ex,partopsep=1ex,parsep=1ex]
\item \textbf{Controller:} Since all invocations pass through the Load
  Balancer, it is the ideal place to manage histograms and other
  metadata required for the hybrid policy.  We add new logic to the
  Load Balancer to implement the hybrid policy and to update the
  keep-alive and pre-warm parameters after each invocation.  We also
  modify the Load Balancer to publish the pre-warming messages.

\item \textbf{API:} We send the latest keep-alive parameter for a
  function to the corresponding Invoker alongside the invocation
  request.  To do this, we add a field to the \emph{ActivationMessage}
  API, specifying the keep-alive duration in minutes.
\item \textbf{Invoker:} The Invoker unloads Docker containers that
  have timed-out in the \emph{ContainerProxy} module.
  We modify this module to unload containers based on the keep-alive
  parameter received from \emph{ActivationMessage}.
\end{enumerate}

\section{Evaluation}

\subsection{Methodology}

\myparagraph{Simulator} Evaluating our policy requires (1) long
executions to assess applications with infrequent invocations, and (2)
exploring a large space of configurations.  To limit the evaluation
time, we use simulations.  We build a simulator that allows us to
compare various policies using real invocation traces.

The simulator generates an array of invocation times for each unique
application.  It then infers whether each invocation would be a cold
start.  By default, the first invocation is always assumed to be a
cold start.  The simulator keeps track of when each application
image is loaded and aggregates the wasted memory time for the
application, \ie the time when the application's image was kept in
memory without actually executing any functions.  We conservatively
simulate function execution times equal to 0 to quantify the
worst-case wasted resource time.  We do not have memory usage data
for all applications, so we also simulate that applications use the
same amount and focus on the wasted memory time.

\myparagraph{Real experiments} To show that our policy can be easily
implemented in real systems with minimal overheads, we use our
OpenWhisk implementation (Section~\ref{sec:implementation}).  Our
setup consists of 19 VMs. One VM with 8 cores and 8GB of memory hosts
containers for the Controller and other main components, including
Nginx and Kafka.
Each of the remaining 18 VMs has 2 cores and 4GB of memory, hosting an
Invoker to actually run the functions in Docker containers.

\myparagraph{Workloads} As input to our simulations, we use the
first week of the trace from Section~\ref{sec:characterization}.
For the real experiments, we use a scaled-down version of the trace.
We randomly select applications with mid-range popularity.  As we
run the full system, we limit each OpenWhisk execution to only 8
hours.  As we show in Section~\ref{sec:experiments}, {\em the
experimental and simulation results show the same trends in both
cold start and memory consumption behaviors.}

\begin{figure}[!t]
    \centering
    \includegraphics[width=0.9\linewidth]{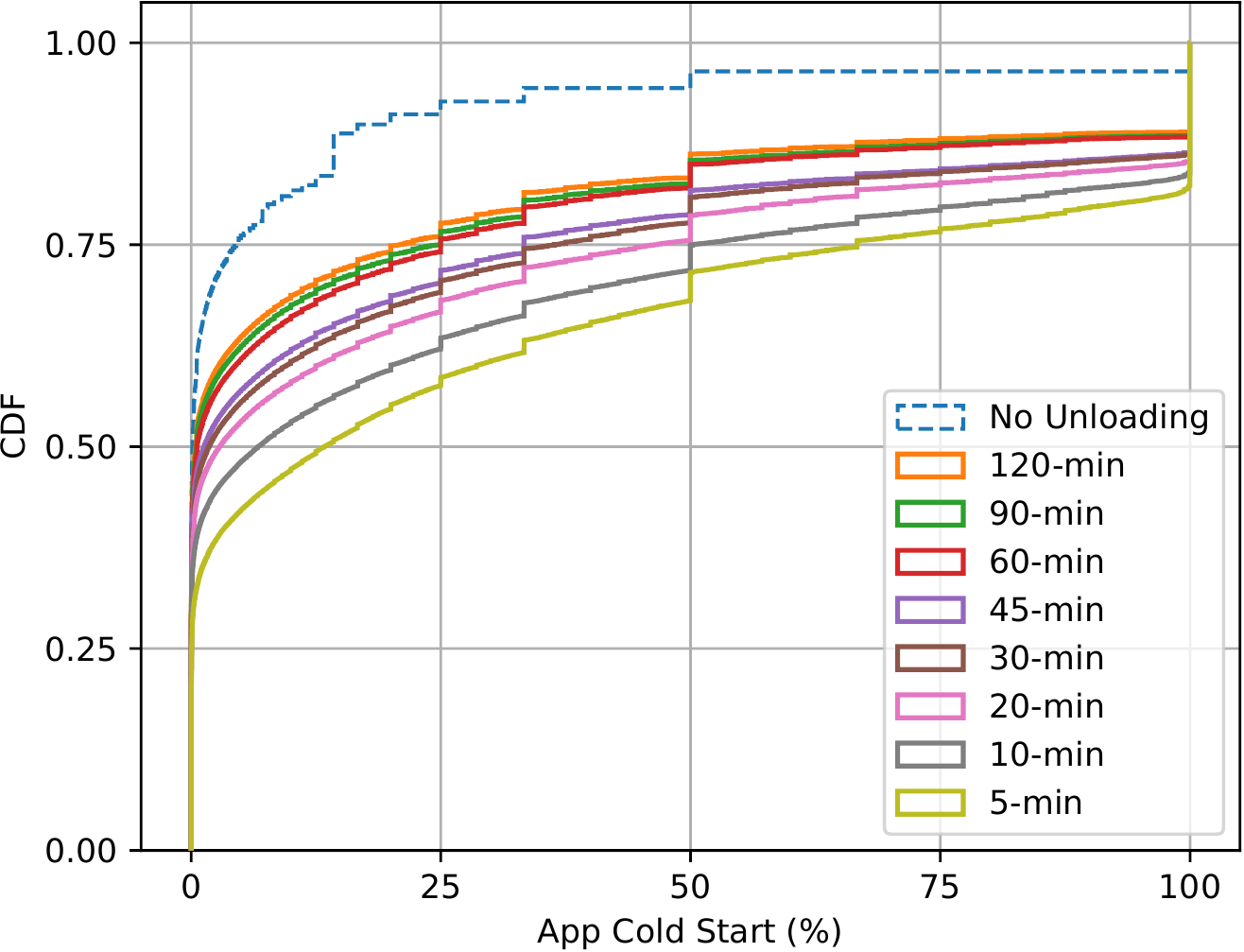}
    \vspace{-.1in}
    \caption{Cold start behavior of the fixed keep-alive policy, as a
      function of the keep-alive length.}
    \label{fig:fixed_cdf_comp}
    \vspace{-.2in}
\end{figure}

\subsection{Simulation Results}

\myparagraph{Understanding the fixed keep-alive policy} We start
evaluating the policy used by most providers: the fixed keep-alive
policy.  We first assess how the length of the keep-alive affects the
cold starts.  Figure~\ref{fig:fixed_cdf_comp} shows the distribution
of cold start percentage experienced by all applications for various
lengths.  For comparison, we also include a \emph{No unloading}
policy, which corresponds to each application only experiencing the
initial cold start.  Even the \emph{No unloading} policy has
$\sim$3.5\% of applications with 100\% cold starts; these applications
have only one invocation in the entire week.

We see significant cold start reduction going from a 10-minute
keep-alive to 1-hour.  The \nth{75}-percentile application experiences
cold starts 50.3\% of the time for the 10-minute keep-alive.  This
number goes down to 25\% for 1-hour.  The cold start improvement is
more pronounced in the last quartile of the distribution, since
applications with infrequent invocations are those that benefit the
most.  From now on, we will focus on this metric (\ie,
\nth{75}-percentile) to report cold starts.

\begin{figure}[!t]
    \centering
    \includegraphics[width=0.9\linewidth]{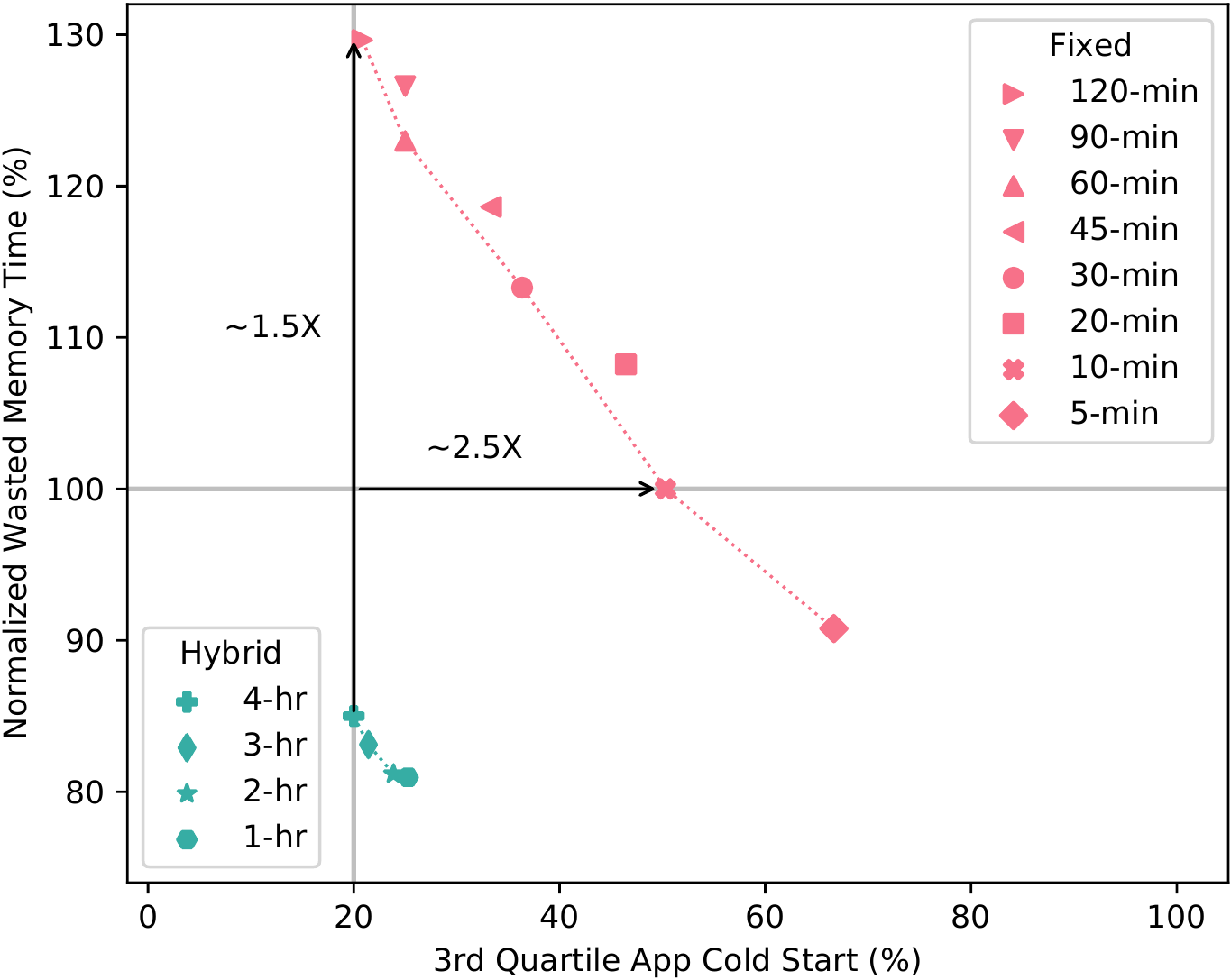}
    \vspace{-.1in}
    \caption{Trade-off between cold starts and wasted memory time for
      the fixed keep-alive policy and our hybrid policy.}
    \label{fig:hist_len_fix_pareto}
    \vspace{-.2in}
\end{figure}

While a longer keep-alive reduces cold starts significantly, it also
increases the resources wasted significantly.  The red markers in
Figure~\ref{fig:hist_len_fix_pareto} show the trade-off between cold
starts and memory wasted, where we normalize the wasted memory time to
the 10-minute keep-alive.  The red curve near the red markers
approximates the Pareto curve.  The figure shows, for example, that a
fixed 2-hour keep-alive has almost 30\% higher wasted memory time than
the 10-minute baseline.
An optimal policy would deliver the lowest cold starts with minimum
cost.  We rely on these Pareto curves to evaluate the policies.

\myparagraph{Impact of using a histogram} We now start to evaluate our
hybrid policy with the impact of the histogram and its range.  The
green markers in Figure~\ref{fig:hist_len_fix_pareto} show the cold
start percentage and wasted memory time of our histogram for various
ranges.  The figure shows how our policy reduces the cold starts
significantly with lower memory waste.  In fact, the 10-minute fixed
keep-alive policy involves $\sim$2.5x more cold starts at the
\nth{75}-percentile while using the same amount of memory as our
histogram with a range of 4 hours.  From a different perspective, the
fixed 2-hour keep-alive policy provides roughly the same percentage of
cold starts as the 4-hour histogram range, but consumes about 50\%
more resources.  Overall, the hybrid policies form a parallel, more
optimal Pareto frontier (green curve) than the fixed policies (red
curve).

\begin{figure}[!t]
    \centering
    \includegraphics[width=\linewidth]{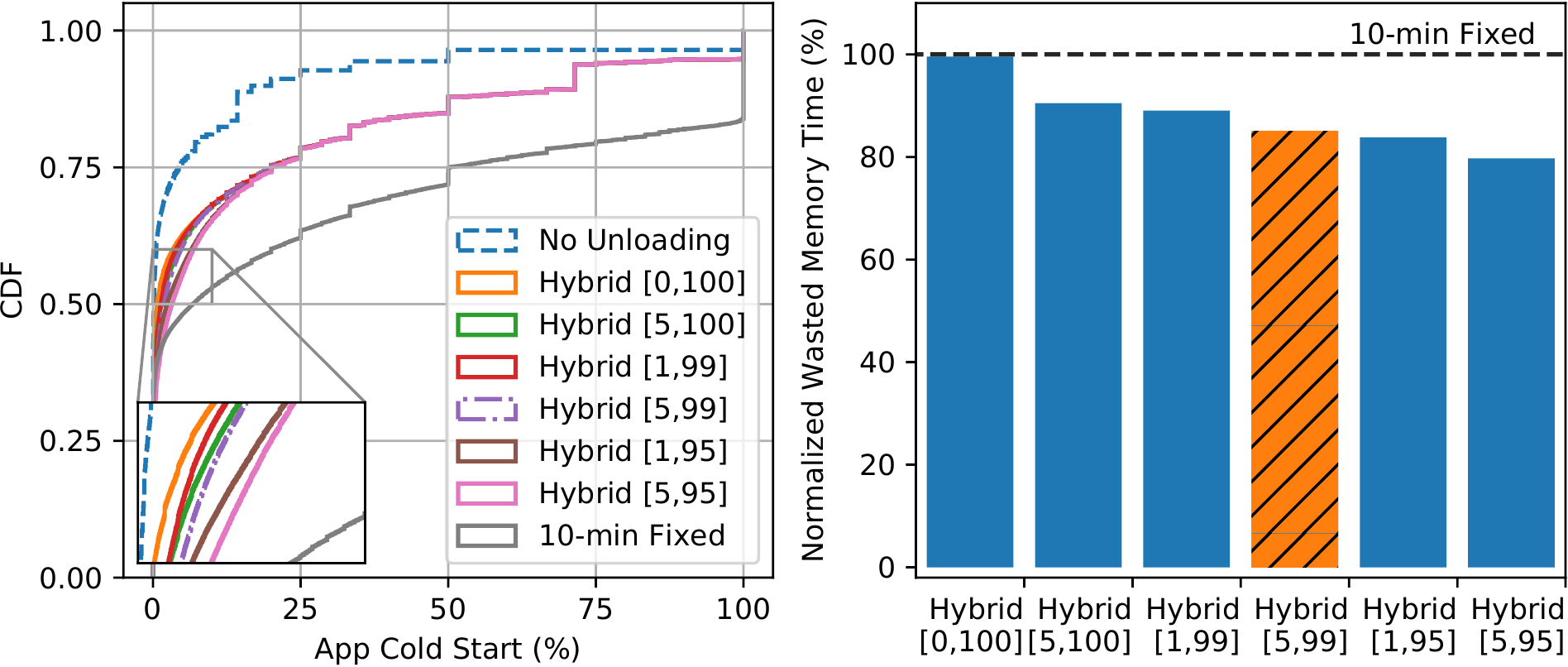}
    \vspace{-.25in}
    \caption{Wasted memory time can be significantly reduced by
      excluding outliers from the IT distribution.}
    \label{fig:cu_percentile_effect}
    \vspace{-.05in}
\end{figure}

\myparagraph{Impact of the histogram cutoff percentiles} Our policy
uses two cutoff percentiles to exclude outliers in the head and tail
of the IT distribution.  Figure~\ref{fig:cu_percentile_effect} shows
the sensitivity study that we used to determine suitable cutoff
values.  The figure shows that, by setting the head and tail cutoffs
to the \nth{5}- and \nth{99}-percentiles of the IT distribution
(labeled {\em Hybrid[5,99]} in the figure), the cold start percentage
does not degrade noticeably whereas the wasted memory time goes down
by 15\%, compared to the case with no cutoff ({\em Hybrid[0,100]}).

\begin{figure}[!t]
    \centering
    \includegraphics[width=\linewidth]{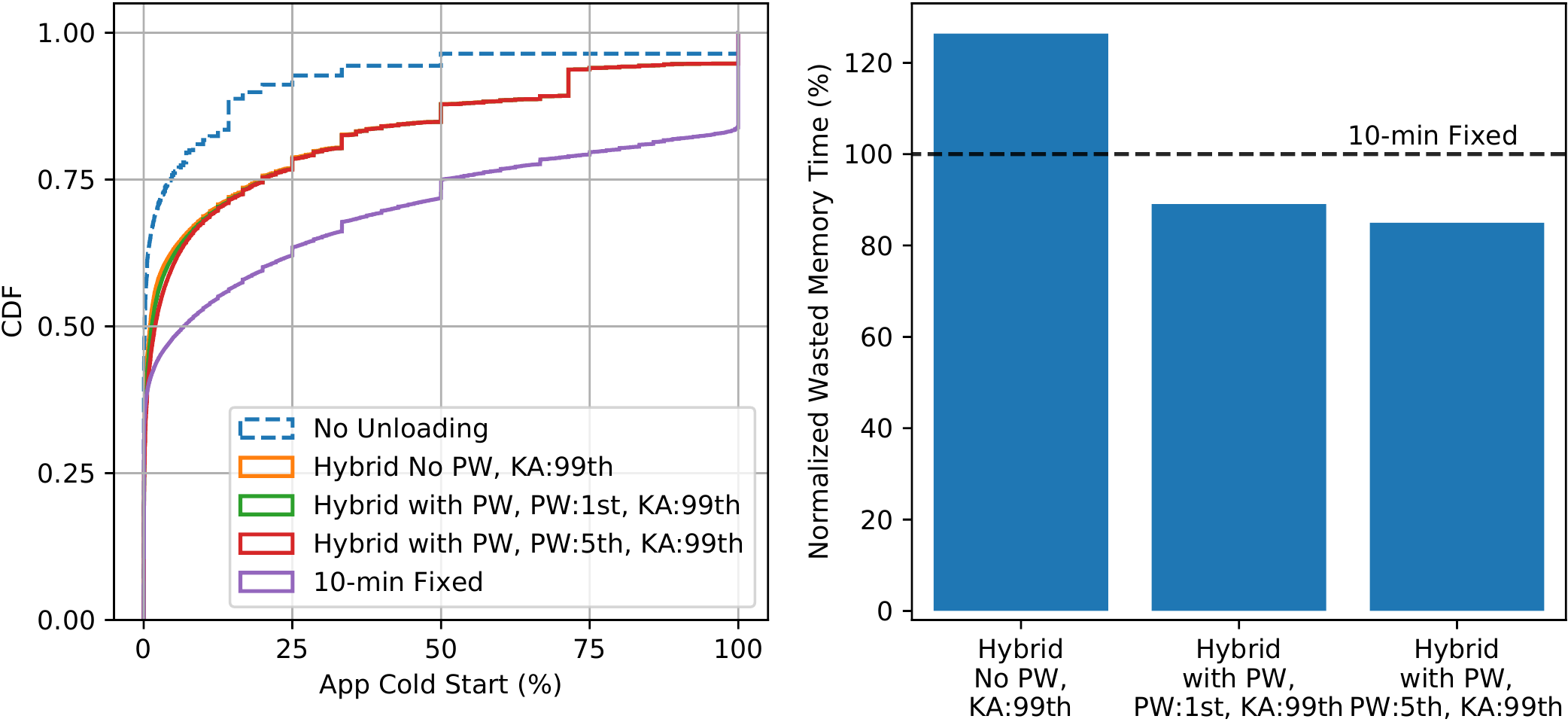}
    \vspace{-.25in}
    \caption{Pre-warming reduces the wasted memory time
      significantly. The cost is slight increase in cold starts.}
    \label{fig:pw_benefit}
    \vspace{-.05in}
\end{figure}

\myparagraph{Impact of unloading and pre-warming} Complementing
  our adaptive keep-alive with pre-warming allows unloading of an
  application right after execution and pre-warming right before the
  next invocation.  This reduces the wasted memory time of application
  images.  Figure~\ref{fig:pw_benefit} shows this, where using similar
  keep-alive (KA) configurations with and without pre-warming (PW) has
  significantly different wasted memory time.  The cost, however, is
  adding a small number of cold starts from unexpected invocations.
  We can control this trade-off by adjusting the histogram head cutoff
  percentile.

\begin{figure}[!t]
    \centering
    \includegraphics[width=1\linewidth]{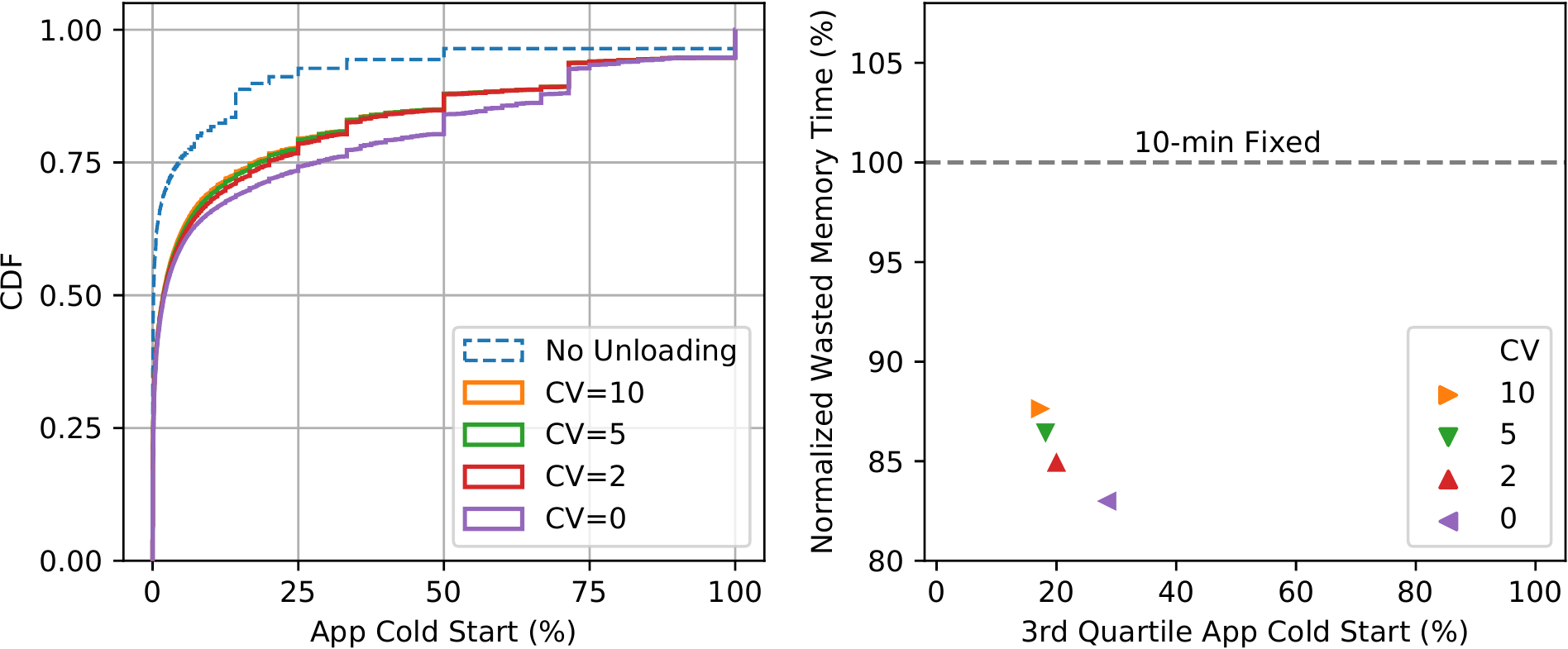}
    \vspace{-.25in}
    \caption{Trade-off between cold starts and memory wasted, as a
      function of the CV threshold, using a 4-hour range.}
    \label{fig:cv_variatoin_cdf_and_pareto}
    \vspace{-.2in}
\end{figure}

\myparagraph{Impact of checking the histogram representativeness} Our
policy checks whether the histogram is representative before using it.
If the histogram is not representative (\ie, the CV of its bin counts
is lower than a threshold), it uses a standard keep-alive approach
where applications stay loaded for the same length as the histogram
range.  We study the impact of different CV thresholds in
Figure~\ref{fig:cv_variatoin_cdf_and_pareto}.  The figure shows the
application cold start distributions (left) and the Pareto frontier
(right).  We see significant gains using a small CV threshold larger
than 0.  We opt for CV=2 as our default threshold.  Increasing the CV
further has negligible cold start reduction with higher resource
costs.

\begin{figure}[!t]
    \centering
    \includegraphics[width=1\linewidth]{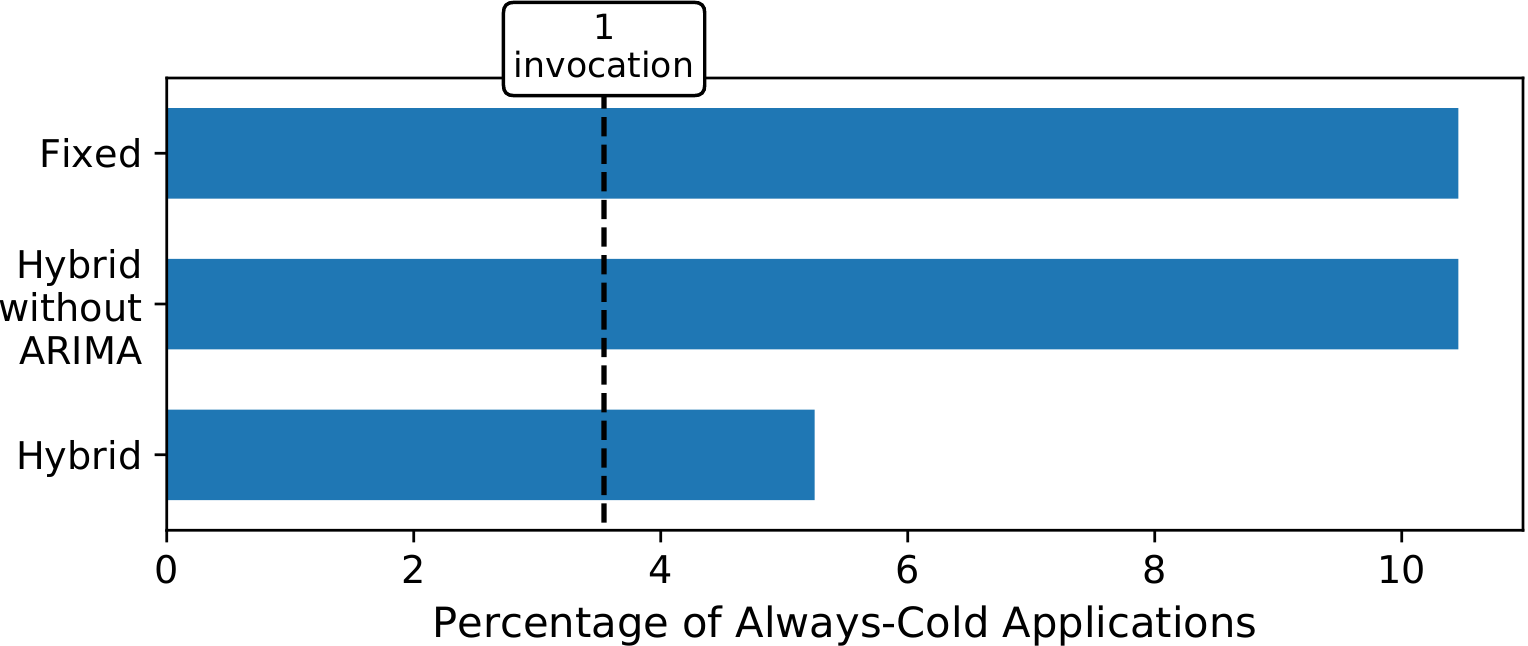}
    \vspace{-.25in}
    \caption{Percentage of applications that always experience cold
      starts, as a function of policy.}
    \label{fig:always_cold_comparison}
    \vspace{-.2in}
\end{figure}

\myparagraph{Impact of using time-series analysis} Another feature
of our hybrid policy is to use ARIMA modeling for applications that
have many ITs outside the range of the histogram. To evaluate its
impact, we now focus on the percentage of applications that show
100\% cold starts.  Figure~\ref{fig:always_cold_comparison} shows
this percentage when using (1) the fixed keep-alive policy, (2) the
hybrid policy without ARIMA, and (3) the full hybrid policy
(including ARIMA).  All of them use 4 hours for the fixed keep-alive
and the histogram range.  During the week-long simulation window,
0.64\% of invocations were handled by ARIMA, and 9.3\% of
applications used ARIMA at least once.  Using ARIMA reduces the
percentage of applications that experience 100\% cold starts by
about 50\%, \ie from 10.5\% to 5.2\% of all applications.  A
significant portion of these applications have only one invocation
during the entire week and no predictive model can help them.
Excluding these applications, the same reduction becomes 75\%, \ie
from 6.9\% to 1.7\% of all applications.  This shows that ARIMA
provides benefits for applications that cannot benefit from a fixed
keep-alive or a histogram-based policy.

\myparagraph{Summary} Our hybrid policy can reduce the number of cold
starts significantly while minimizing the memory cost.  We achieve
these positive results despite having deliberately designed our policy
for simplicity and practicality: (1) histogram bins have a resolution
of 1-minute, (2) histograms have a maximum range, (3) they do not
require any pre-processing or complicated model updates, and (4) when
the histogram does not work well, we resort to simple and effective
alternatives.

\subsection{Experimental results}
\label{sec:experiments}

We ran two experiments with 68 randomly selected mid-range popularity
applications from our workload on our 19-VM OpenWhisk deployment: one
experiment with the default 10-minute fixed keep-alive policy of
OpenWhisk, and another with our hybrid policy and a 4-hour histogram
range.  Each experiment ran for 8 hours.  During the 8-hour period,
there are a total of 12,383 function invocations.  We use
FaaSProfiler~\cite{FaaSProfiler_2020,Shahrad_2019} to automate trace
replay and result analysis.

\begin{figure}[!t]
    \centering
    \includegraphics[width=0.8\linewidth,height=1.8in]{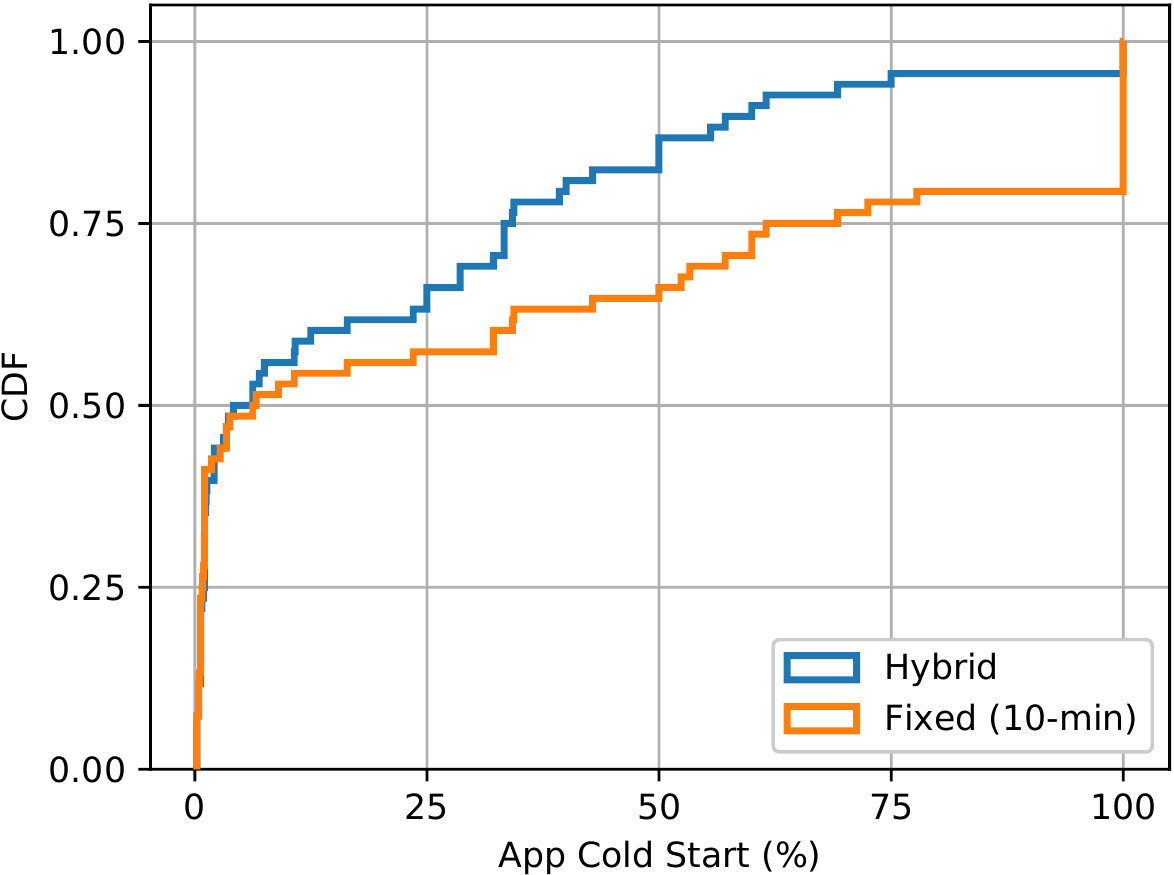}
    \vspace{-.1in}
    \caption{Cold start behavior of fixed keep-alive and hybrid
      policies in OpenWhisk.}
    \label{fig:experimental_cdf}
    \vspace{-.2in}
\end{figure}

Figure~\ref{fig:experimental_cdf} compares the cold start behavior of
the hybrid and 10-minute fixed keep-alive policies.  The significant
cold start reductions follow the {\em same trend as our simulations}
(left graph of Figure~\ref{fig:cu_percentile_effect}).
On average and across the 18 Invoker VMs, the hybrid policy reduced
memory consumption of worker containers by 15.6\%, which is also {\em
  consistent with our simulation results} (right graph of
Figure~\ref{fig:cu_percentile_effect}).  Moreover, the hybrid policy
reduced the average and 99-percentile function execution time 32.5\%
and 82.4\%, respectively.  This is due to a secondary effect in
OpenWhisk, where the language runtime bootstrap time is eliminated for
warm containers.

\myparagraph{Policy overhead} 
We measure the (1) additional latency induced by our implementation
and (2) the impact of our policy on the scalability of the OpenWhisk
controller.  The Scala code that implements our policy in the
Controller adds an average of only 835.7$\mu s$ ($\sigma=245.5\mu s$)
to the end-to-end latency.  This overhead is negligible compared to
the existing latency of OpenWhisk components: the (in-memory) language
runtime initiation takes $O(10ms)$ and the container initiation takes
$O(100ms)$ for cold containers~\cite{Shahrad_2019}.  For the uncommon
cases where ARIMA is required (0.7\% of invocations), the initial
forecast involves building the model, which takes an average of
26.9ms, whereas subsequent forecasts take an average of 5.3ms.  Since
ARIMA works for applications that would normally experience cold
starts, these overheads represent a relatively small cost compared to
the cold start overhead.

In terms of scalability, CPU utilization is the limiting factor for
the Controller.  Our policy adds only a 4-6\% higher utilization for a
range of benchmarking request rates (10rps to 300rps), compared to
OpenWhisk's default policy.

\section{Production Implementation}
We have implemented our policy in Azure Functions for HTTP-triggered
applications; its main elements will be rolling out to production in
stages starting this month.  Here, we overview the implementation.

Azure Functions has a controller that communicates with
function-execution workers through HTTP, and a database for
persisting system state.  The controller gets asynchronous updates
from the workers at fixed intervals; we use these to populate the
histogram.  We keep the histogram in memory (bucket of 240 integers
per application, or 960 bytes) and do hourly backups to the
database.  We start a new histogram per day in the database so we
can track changes in application’s invocation pattern, and remove
histograms older than 2 weeks.  We can potentially use these daily
histograms in a weighted fashion to give more importance to recent
records.
 
When an application changes state from executing to idle, we use the
aggregated histogram to compute its pre-warm interval and schedule
an event for that time (minus 90 seconds).  Pre-warming loads
function dependencies and performs JIT where applicable. Some steps,
like JIT of the function code, happen when the actual invocation
comes in as the function’s code cannot be executed as part of warmup
to preserve execution semantics. Each worker maintains the
keep-alive duration separately, depending on how long it has been
idle for.  We make all policy decisions asynchronously, off the
critical path to minimize the latency impact on the invocation. This
includes updating the in-memory histogram, backing up the histogram
to the database, scheduling pre-warming events, and controlling the
workers' keep alive intervals.

\section{Related Work}
\label{sec:related}

There is a fast-increasing number of studies on different aspects
of serverless computing. The most relevant for our paper are those
that characterize FaaS platforms and applications, and those that
propose and optimize FaaS serving systems.

\myparagraph{FaaS characterization} A few studies~\cite{Figiela_2018,Kuhlenkamp_sac_2020, Lee_2018,back2018using,lloyd2018serverless,Wang_2018} 
have characterized the main commercial FaaS providers, {\em but only from
  the perspective of external users.}
They typically reverse-engineer aspects of FaaS offerings, by running
benchmark functions to collect various externally visible metrics.
Our characterization is orthogonal to these works, as we provide a
longitudinal characterization of the entire workload of a large cloud
provider from the provider's perspective.  {\em Our characterization
  is the first of its kind.}

Another class of studies looks at the ways developers are using FaaS
offerings, by looking at public application
repositories~\cite{Spillner_2019}. While valuable, this approach
cannot offer insights on the aggregate workload seen by a provider.

\myparagraph{Optimizing FaaS serving} Another set of relevant work
considers optimizing different aspects of FaaS systems.  Van Eyk
\etal~\cite{van2018spec} identify performance-related challenges,
including scheduling policies that minimize cold starts. They also
identify the lack of execution traces from real FaaS platforms as a
major obstacle to addressing the challenges they identified.

For optimizing each cold start, Mohan \etal~\cite{mohan2019agile}
find that pre-allocating virtual network interfaces that are later
bound to new function containers can significantly reduce cold start
times.  SOCK~\cite{oakes18sock} proposes to optimize the loading of
Python functions in OpenLambda by smart caching of sets of
libraries, and by using lightweight isolation mechanisms for
functions.  SAND~\cite{akkus2018sand} uses application-level
sandboxing to prevent the cold start latency for subsequent function
invocations within an application. Azure Functions warms all
functions within an application together; thus this was not a
concern for us.  Replayable Execution~\cite{wang2019replayable}
proposes checkpointing and sharing of memory among containers to
speed up the startup times of a JVM-based FaaS system.  Kaffes
\etal~\cite{Kaffes_2019_socc} propose a centralized core-granular
scheduler.  {\em Our work on reducing the number of cold starts and
resource usage by predicting function invocations is orthogonal to
these improvements.}

Other studies also use prediction to optimize different aspects.
Work in~\cite{hoseinyfarahabady2017dynamic,
hoseinyfarahabady2017model} proposes a policy for deciding on
function multi-tenancy, based on a predictive model of resource
demands of each function.  Without discussing design details,
EMARS~\cite{saha2018emars} proposes using predictive modeling for
allocation of memory to serverless functions.
Kesidis~\cite{kesidis2019temporal} proposes to use the prediction of
the resource demands of functions to enable the provider to overbook
functions on containers.  In contrast, we track invocation patterns
and use this knowledge to reduce cold starts and memory waste. 

\myparagraph{Cache management} 
Finally, one might think that the problem of managing cold starts is
similar to managing caches of variable-sized objects, such as Web
page caches and
others~\cite{balamash2004overview,Ali11,Podlipnig03}.  However,
there are two fundamental differences.  First, FaaS frameworks are
often implemented on top of services that charge by the time
resources are allocated (\eg, each application is packaged as a
container and deployed to a container service).  Thus, cold start
policies proactively unload applications/functions from memory,
instead of waiting for other applications/functions to need the
space. Our policy is closest to a class of TTL-based caches where
new accesses reset the TTL~\cite{basu18ttl,berger2014ttl}.  These
works did not consider temporal prefetching, the equivalent of our
pre-warming.  Other caching work did consider it, but with
capacity-based replacements~\cite{wu19prefetching}.  Second, most
caching algorithms to date have focused on aggregate performance
metrics~\cite{dehghan2019utility,ferragut2016optimizing}, such as
the weighted sum or average of per-object miss ratios. In contrast,
we tailor our cold start management to each application to maximize
individual customer satisfaction.

\section{Conclusion}

In this paper, we characterized the entire production FaaS workload
of Azure Functions.  The characterization unearthed several key
observations for cold start and resource management.  Based on them,
we proposed a practical policy for reducing the number of cold
starts at a low resource cost.  We evaluated the policy using both
simulations and a real implementation, and real workload traces.
Our results showed that the policy can achieve the same number of
cold starts at much lower resource cost, or keep the same resource
cost but reduce the number of cold starts significantly. Finally, we
overviewed our policy's implementation in Azure Functions.  We
released sanitized traces from our characterization data
at~\cite{Trace}.  \subsection*{Acknowledgements} We would like to
thank our shepherd, George Amvrosiadis, and the anonymous reviewers
for helping us improve this paper.  We also thank Daniel Berger,
Bill Bolosky, and Willy Zwaenepoel for their comments on earlier
versions of it.
 

\end{document}